%% file: ofc27y.tex
\documentclass[letter,preprint,aps,longbibliography]{revtex4-1}
\include{header}

\usepackage{xcolor}
\usepackage{multirow}
\usepackage{amsmath}
\usepackage{graphicx}
\usepackage{epsfig}
\usepackage{simplewick}
\usepackage{caption}
\usepackage[colorlinks = true]{hyperref}
\newlength{\upit}\upit=0.1truein

\newcommand{\ltappr}{{{\lower4pt\hbox{$<$} } \atop \widetilde{ \ \ \ }}}
\newlength{\bxwidth}\bxwidth=1.5 truein

\newcommand{\Ua}{\Uparrow}
\newcommand{\Da}{\Downarrow}
\newcommand{\up}{\uparrow}
\newcommand{\dw}{\downarrow}
\newcommand{\dsp}{\displaystyle}

\renewcommand\joinrel{\mathrel{\mkern-9mu}}

\newcommand\relbd{\mathrel{{\bf\smash{{\phantom- \above1pt \phantom-
}}}}}
\newcommand\ltdash{\raise-0.7pt\hbox{$\scriptscriptstyle |$}}

\newlength{\figwidth}
\newlength{\shift}
\newcommand{\fg}[3]
{
\begin{figure}[ht]

\vspace*{-0cm}
\[
\includegraphics[width=\figwidth]{#1}
\]
\vskip -0.2cm

\caption{\label{#2}
\small#3
}
\end{figure}}

\newcommand{\bk}{{\bf{k}}}

\graphicspath{{Figures/}}
\begin{document}
\title{Order Fractionalization
}

\author{Yashar Komijani$^{1}$, Anna Toth$^{2}$, Premala Chandra$^{1}$
and Piers Coleman$^{1,3}$}
\affiliation{
$^{1}$Center for Materials Theory, Department of Physics and Astronomy,
Rutgers University, 136 Frelinghuysen Rd., Piscataway, NJ 08854-8019, USA}
\affiliation{$^{2}$ Bajza utca 50., H-1062 Budapest, Hungary}
\affiliation{$^{3}$ Department of Physics, Royal Holloway, University
of London, Egham, Surrey TW20 0EX, UK.}
\date{\today}
\begin{abstract}
{\bf 
The confluence of quantum mechanics and complexity, which
leads to the emergence of 
rich, exotic states of matter, motivates the extension of  our
concepts of quantum ordering. The twin concepts of spontaneously 
broken symmetry,  described in terms of a Landau order parameter, 
and of off-diagonal
long-range order (ODLRO),
are
fundamental to our understanding of phases of matter.
In electronic matter it has long
been assumed that Landau order parameters
involve an even number of electron fields,
with  integer spin and
even charge, that are bosons.  
On the other hand, in low-dimensional magnetism, operators are known to
fractionalize so that the excitations carry spin-1/2.
Motivated by experiment, mean-field theory and computational results,
we extend the concept of
ODLRO into the time domain, proposing that in a broken symmetry state,  
quantum operators can fractionalize into half-integer order parameters.
Using numerical renormalization group studies we show how such
fractionalized order can be  
induced in quantum impurity models. We then
conjecture that 
such order develops 
spontaneously in lattice quantum systems, due to positive feedback,
leading to a new family of phases, manifested by a coincidence of
broken symmetry and fractionalized excitations that can be detected by
experiment.

}
\end{abstract}

%
\maketitle


A major theme in current condensed matter physics is the quest for new 
types of quantum matter such as high-temperature superconductors, 
topological insulators and spin liquids \cite{everyone2006,palee06,qimiao16,jemoore,sczhang11,kanehasan,balents17,kanoda}.  
An important aspect of this research is the
characterization of novel forms of order that 
emerge in these quantum materials; another concerns the new classes
of excitation that accompany these orderings. 
Landau's theory of phase transitions  \cite{landaupt}
attributes the transformation in macroscopic properties to the
development of an order parameter that breaks the microscopic
symmetries of the system. 
Later, Yang observed \cite{Yang62} 
that such long-range order 
is manifested as an asymptotic factorization
of spatial correlation functions at long distances 
into a product of order parameters 
$
\langle  {\cal O}(x){\cal O}\dg (y)\rangle \xrightarrow{|x-y|\rightarrow
\infty }\langle  {\cal O} (x)\rangle  \langle  {\cal  O}(y)\rangle^{*}
$.
The quantum operators ${\cal  O}$ are bosonic and 
condense into a
state of ``Off-Diagonal Long Range Order'' (ODLRO).


In relativistic physics half-integer spin order parameters are
prohibited by the spin-statistics theorem  \cite{pauli40,Duck:1998be}, but in 
electronic condensed matter the absence of Lorentz invariance removes this
restriction. Though half-integer order
can be envisioned in Landau's theory of phase
transitions, it is microscopically incompatible with
ODRLO where the local operators that condense are bosons, formed 
from an even number of half-integer spin fermions.
Conventional order parameters such as magnetization or 
pair density involve pairs of fermions and form part of the general
paradigm of BCS/Hartree-Fock order parameters; 
more complicated ``composite
order parameters'' involving four or more elementary fermion fields
have also been envisioned  \cite{emery92,bonca93}, but all have integer spin.  This has led to
the implicit assumption that in electronic quantum matter, order
parameters satisfy an effective spin-statistics theorem, carrying 
integer spin and even charge. 

Another important development in condensed matter physics is the discovery of
``fractionalization'', where the emergent excitations 
carry fractional quantum numbers\cite{Wilczek82,fradkin,laughlin83,Anderson:1987ii}.
A classic  example is the one dimensional spin-$1/2$ Heisenberg
antiferromagnet where a spin-flip, that changes the magnetic
quantum number by an integer unit,  creates a pair
of spin-$1/2$ excitations called spinons   \cite{tenant93,mourigal13}.
Higher dimensional examples include the fractional quantum Hall
effect \cite{laughlin83}, and spin liquids like the Kitaev
honeycomb model where the spin operators fractionalize into
Majorana fermions \cite{Kitaev06}. 
Fractionalization has also been proposed to occur  
at continuous 
quantum  phase transitions  \cite{Senthil04,Sachdev11} leading to 
``deconfined quantum criticality''
where a fluctuating order parameter
breaks up into new degrees of freedom. 
%
%
Whereas ODLRO is a ground-state property, fractionalization is associated with
excitations, manifested in dynamical response functions and as 
correlations that are nonlocal in time.
In this paper we explore the possible unification of ODLRO and 
fractionalization, 
proposing that quantum operators can 
fractionalize into half-integer order parameters. 
This order fractionalization conjecture (OFC)
requires an extension of ODLRO into space-time, and 
suggests a new symmetry class of {quantum order}  \cite{Wen02}.


A key setting for our discussion is
the Kondo lattice, a model describing an array of magnetic moments
interacting via an antiferromagnetic exchange with a sea of 
conduction electrons. This model is widely used to describe
the behavior of heavy fermion materials, where the screening 
of the magnetic moments by conduction electrons at low temperatures
liberates their spins into the Fermi sea as delocalized heavy
electrons
(Fig. \ref{fig1}{\bf A}, {\bf B}), a process that enlarges the Fermi surface.
Since a spin flip of a local moment 
creates a particle-hole pair of heavy fermions, 
we are led to interpret the expansion of the Fermi surface
as a fractionalization of local moments
into negatively charged fermions  \cite{Lebanon:2007ip}.
The origin of the moments is immaterial and their
fractionalization into heavy fermions 
would even occur if they were of nuclear origin \cite{coleman2016}. 
(Fig. \ref{fig1} {\bf C}).
\figwidth=\textwidth
\fg{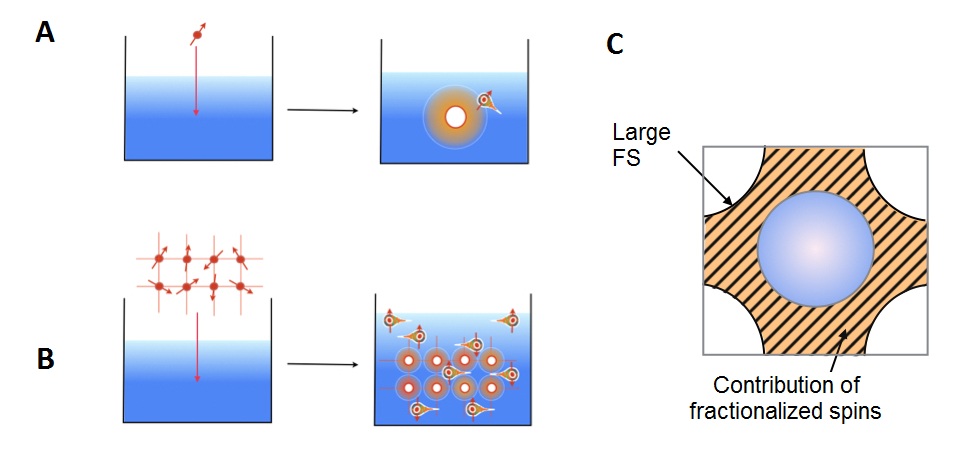}{fig1}
{Schematic illustrating the Kondo effect 
showing {\bf(A)} the fractionalization of a single spin into a delocalized
f-electron in 
a Kondo impurity model, {\bf (B)} the fractionalization of local
moments in a Kondo lattice, to form a fluid of heavy fermions;  {\bf (C) }
the enlargement of the Fermi surface from small (blue) to large (hatched) due to the
formation of heavy fermions, as predicted by 
Oshikawa's theorem \cite{Oshikawa00}. }

There is considerable indirect experimental and theoretical support for spin
fractionalization in the Kondo lattice.
Using topology, Oshikawa has shown that in a Fermi liquid ground-state, 
the screened spins of a Kondo lattice
contribute to an expansion of its Fermi surface volume \cite{Oshikawa00}. 
Hall effect and de Haas van Alphen measurements subsequently detected jumps
in the Fermi surface volumes  at quantum phase transitions between
antiferromagnetic and paramagnetic heavy fermion 
ground-states \cite{Paschen04,Shishido05}. 
The enlargement of the Fermi surface in
the Kondo lattice indicates the formation of
half-integer excitations from the lattice of local moments, 
a process that is most naturally interpreted as spin fractionalization. 


Experimentally, there are important examples where Kondo spin
screening appears coincident with the development of long range order.
For instance, in both  NpPd$_{5}$Al$_{2}$ and CeCoIn$_{5}$,
singlet superconductivity develops directly from a Curie-Weiss
paramagnet \cite{aoki,sarrao} with a substantial loss of spin entropy.
Similar phenomenon involving quadrupole degrees of freedom 
has been proposed for UBe$_{13}$, URu$_{2}$Si$_{2}$ and Pr
X$_{2}$Al$_{20}$ (X=Ti, Va)  \cite{Jarrell96,Chandra2013,Dyke18}.
Theoretical evidence for spin fractionalization and broken symmetry 
is found from path-integral based, large-$N$
treatments of Kondo lattices
 \cite{read83_1,Coleman94,Coleman:1999vb,Flint:2008fk}. However, while these
methods demonstrate the feasibility of order
fractionalization in models with very large numbers of spin components, they are unable to demonstrate that this phenomenon extends
to physical, spin-$1/2$ Kondo lattices.
Our motivation to seek new classes of broken symmetry derives
from these experimental and theoretical considerations.


Developing this idea, we recall that 
the dynamics of an
interacting fermion is determined by the Dyson self-energy, 
$\Sigma_{\alpha \beta } (2,1)$, an amplitude
for the scattering of a single-particle excitation, a fermion,  at
space time $1= (x_{1},t_{1})$ back into a single-particle state at the space-time $2\equiv
(x_{2},t_{2})$ via intermediate many-body states.  Here $\beta $ and $\alpha$ are the 
internal quantum
numbers of the incoming and outgoing fermions. The Hamiltonian
that determines the time evolution is invariant under various 
global symmetry transformations such as spin rotation or global gauge
invariance; at high temperatures the self-energy is also invariant 
under these symmetries.   However if a phase transition occurs, 
the self-energy develops a symmetry-breaking component 
resulting from scattering off the order parameter.
For example a ferromagnet develops a 
spontaneous Zeeman splitting driven by the internal Weiss field;
a BCS superconductor develops a pairing field due 
to Andreev scattering off the condensate. 
{In all these classical examples, the order parameter has an associated coherence length; for space/times larger than this coherence length, the (coarse-grained) self-energy can be regarded as a local, instantaneous symmetry-breaking potential}, $\Sigma_{\alpha \beta } (2,1)=
M_{\alpha \beta } (1)\delta (2-1)$, where the order parameter 
$M_{\alpha \beta } (1)$ transforms
as an irreducible representation of the Hamiltonian 
symmetry group (Fig. \ref{fig2}A). 

Fractionalization implies a factorization of quantum
operators into two or more independent components.
Similarly, we take order fractionalization to imply
that at large space-time separations between $1$ and $2$, 
the self-energy factorizes into a 
product of fractional order parameters, 
$\Sigma_{\alpha \beta }
(2,1)\sim  \bar V_{\alpha } (2) V_{\beta } (1)$, where
$ V_{\beta }(1)$
and 
$\bar V_{\alpha } (2)$ 
 describe a fractional,
spinorial order parameter and its conjugate at locations $1$ and $2$,
respectively. 
The independence of these two quantities 
requires that the intermediate 
state, which involves an odd number of fermions,
develops a bound-state that propagates without decay between $1$ and
$2$,  as 
shown schematically in Fig. \ref{fig2}{\bf B}.   
Since $V_{\beta}$ carries the quantum number of the incoming
fermion, 
the intermediate bound-state fermion is neutral with respect to this
quantum number. 
This establishes a link between order fractionalization and the formation of neutral fermion bound-states.  


Following the example of the Curie-Weiss
theory of magnetism, here we develop support for the 
order fractionalization
conjecture. There, the first step is to
induce Curie magnetism with an external magnetic field acting on a
single spin; next, one argues that in the bulk the 
interaction of one site on another provides a Weiss field that maintains 
a spontaneous magnetization \cite{Weiss1907}. 
Similarly here we seek to {\it induce} order fractionalization
in Kondo impurity models by identifying an appropriate symmetry breaking
field. This is a necessary pre-condition for us to argue 
that ``fractionalizing Weiss fields'' can spontaneously stabilize 
order fractionalization in lattices. 

\noindent {\bf Spin 
Fractionalization.}  We first identify spin 
fractionalization in the single impurity Kondo model. Then we show 
we can induce order fractionalization in the two-channel 
Kondo impurity model where
the local moment is screened by two separate conduction channels. 
The impurity Kondo model 
involves an
antiferromagnetic spin exchange interaction between a local
moment and the spin density of the 
conduction sea $H_{I} = J (\psi \dg \vec{\sigma }\psi )\cdot \vec{S}$,
where $\psi \dg$ {is a spinor that} creates an electron at the site of the impurity. 
At low energies this model is a resonant level 
Fermi liquid where the interaction
matrix elements behave as an effective {(renormalized)} Anderson model, $J \psi \dg (\vec{\sigma }\cdot
\vec{S})\psi\rightarrow  (\bar V\psi \dg  f+ V f\dg \psi )$.  The
equivalence of the effective Hamiltonian with its microscopic form 
implies that the operator combination of a spin and a conduction
electron acts as a single, bound-state fermion
\begin{equation}\label{2}
J(\vec{\sigma }_{\alpha \beta }\cdot
\contraction{}
{\vec{S}}
{)}
{\psi }
\vec{S})
\psi_{\beta } = \bar  V \hat f_{\alpha }.
\end{equation}
Here the horizontal line contracting the spin and the fermion implies
that at long times, this 
combination acts as a single composite 
fermion. 
Although this process has been amply demonstrated in large-$N$
calculations \cite{read83_1,onthebrink} and is implicitly guaranteed
by the low energy equivalence of the Anderson and Kondo impurities
models, we now present an explicit demonstration of its occurance in the
spin-1/2 Kondo model using numerical renormalization {group (NRG)} 
methods \cite{Bulla08}.

In NRG the conduction bath is discretized logarithmically, mapping the
model to an impurity spin coupled to a tight-binding
(Wilson) chain with exponentially decaying tunneling amplitude. This
produces the imaginary part of the Green's function at a set of
discrete frequencies, which are then interpolated to produce a 
continuous, analytic function satisfying the necessary sum rules. We then transform the T-matrix of the
conduction electrons $T(z)$ so obtained, to
the irreducible self-energy $\Sigma(z)$ using the relation
\begin{equation}
\Sigma(z)=\frac{T(z)}{1+ g(z)T(z)},
\end{equation}
where $g(z)$ is the bare local Green's function of the conduction
electrons at the position of the impurity.  The unitary 
single-particle scattering
generated by a Kondo singlet at low energies implies that $g (0)T
(0)=-1$ at the Fermi energy and hence a singular structure in
$\Sigma (z\sim 0)$, making 
the extraction of $\Sigma(z)$ sensitive to interpolation errors in 
the NRG. We employ a limiting procedure in which $g (z)\rightarrow
(1-\epsilon)g (z)$ , keeping $\epsilon$ larger
than the interpolation errors induced in $T (z)$ (c.f. Supplementary
Materials A).
\figwidth=0.8\textwidth
\fg{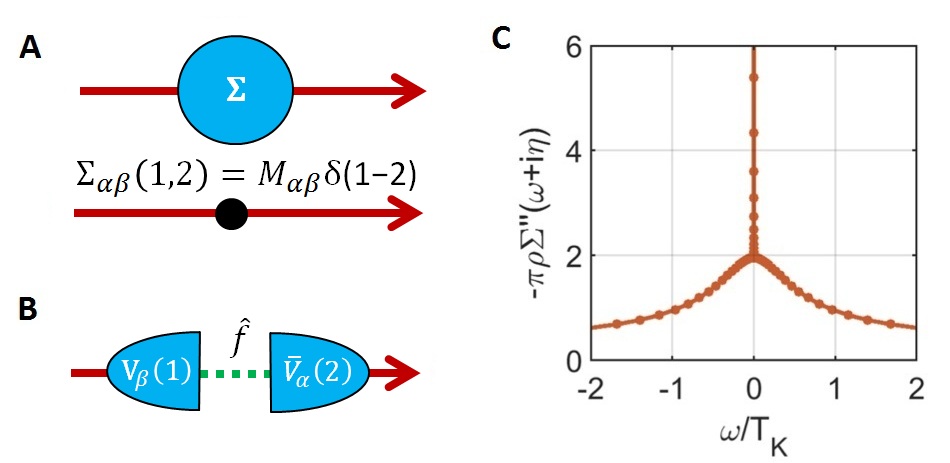}{fig2}{
{\bf(A)} In a conventional broken symmetry state, the coarse-grained
electronic self-energy (top) is
instantaneous and local.  {\bf (B)} Order fractionalization
leads to a factorization of the self-energy into 
two spinorial components, 
linked by a low energy fermionic bound-state 
(bottom). {\bf (C)} The irreducible self-energy in the single-channel
single-impurity Kondo model, computed using NRG, displaying 
a sharp
fermionic pole, on top of a Fermi liquid background (see supplementary
material B).}

Fig. \ref{fig2}{\bf C} shows the result of a NRG   
 calculation of the one-particle irreducible electron self-energy in
 the spin-1/2 Kondo model, 
 indicating that it contains a sharp, resolution-limited pole at zero 
energy, with the asymptotic form
 $\Sigma_{\alpha \beta }\sim \delta_{\alpha\beta }\bar V V/\omega$.   
The pole 
demonstrates the development of a  many-body fermionic bound-state at the Fermi energy
carrying $S=1/2$ and charge $e$; 
the sharpness of the pole confirms
that the spectral decomposition of the 
emergent f-electron field has no overlap with one-particle excitations
of the conduction sea, i.e it is an emergent fermion, described by
the Lagrangian ${\cal L}_{f} = f\dg_{\sigma } (-i\partial_{t})f_{\sigma }
$. The background to the peak can be fit with $\omega^2$ at low energies and is due to Fermi-liquid interactions (c.f. Supplementary Materials B).

To confirm that the local moment fractionalizes into a pair of fermions, we apply a small external magnetic field, under which the 
Lagrangian acquires a term proportional to the magnetization $\vec{M}=\frac{1}{2}\psi\dg\vec\sigma\psi+\vec S$, i.e. 
${\cal L}\rightarrow {\cal L}- \vec{ B}.\vec{M}$. Equivalently, we can
employ a Gallilean transformation 
into a  reference frame rotating with angular velocity
$\vec{\omega}= \left({g\mu_{B}}  \right)\vec{B}$. In the rotating
reference frame, 
$\psi \rightarrow U
\psi $  and $f\rightarrow Uf$ where $U=e^{-i t
\vec{\omega}\cdot\vec\sigma/2}
$, and under this transformation,  
\begin{equation}\label{}
{\cal L}_{f}\rightarrow f\dg
\left(-i\partial_{t} - \vec{\omega}\cdot {\vec{\sigma }}/{2} \right)
f = {\cal L}_{f}- (g\mu_{B})\vec{B}\cdot \vec{S}_{f}
\end{equation}
where $\vec{S}_{f}\equiv  f\dg_{\alpha }
\left(\frac{\vec{\sigma }}{2} \right)_{\alpha \beta }f_{\beta }$. Comparing this to the original $\vec M$, we can identify $\vec S_f$ as the spin, 
fractionalized
into a product of Dirac (i.e. complex) fermions.
%

Traditional treatments of the single channel Kondo model describe the
low energy physics in terms the formation of a Kondo singlet between
an electron and local moment, 
that develops as the Kondo coupling $J$ scales to infinitely strong coupling;
the expulsion of the electrons from the site of the Kondo singlet 
gives rise to unitary scattering, and the development of 
a local Fermi liquid \cite{Nozieres80}. 
To understand how the traditional viewpoint is consistent 
with fractionalization we have analytically
calculated the electron self-energy in the strong coupling limit of large 
$J$, using the Lehmann representation of the
electron Green's function in terms of the exact eigenvalues. 
Remarkably, as in the NRG calculation, the self-energy 
$\Sigma (\omega)= (3J/2)^{2}/{\omega}$ contains a single pole
(Supplementary material D).  Whereas 
we might have expected a large static self-energy of order 
$\Sigma \sim -J$, corresponding to the binding of the
electron into the singlet, instead the Kondo scattering off the singlet
is dynamical. As in the NRG calculation, we are forced to interpret 
the zero-energy pole in the self-energy as the fractionalization
of the local moment into a fermion, hybridized with the conduction sea
to form a local Fermi liquid, as summarized in Fig. \ref{fig1}{\bf A}.
Once we acknowledge that 
scattering off the Kondo singlet is
dynamical at the strong-coupling fixed point, 
we are able to reconcile fractionalization with the
traditional view of the Kondo effect. 

An important aspect of fractionalization is the emergence of an
internal gauge symmetry. The established view is that 
fractionalized excitations carry an internal gauge
charge \cite{Senthil00,
Senthil04,Hansson04,Vishwanath04,Senthil05,Fradkin79}.
The fractionalized spin is invariant under gauge
transformations of the emergent f-excitations,  $f_{\alpha }\rightarrow e^{i\theta
(t)}f_{\alpha }$. Moreover the composite fermion $(\vec{ \sigma
}\cdot \vec{ S})\psi $ involves
a product of the
hybridization and the f-electron, $\bar Vf_{\sigma }$ that is invariant under $U (1)$ transformations of both fields
$f_{\sigma}\rightarrow e^{i\theta (t) }f_{\sigma }$, $V\rightarrow
e^{i\theta (t) }V$.  
When these transformed fields are substituted into the action,
it becomes ${\cal L}_{f}\rightarrow f\dg (i\partial_{t}+A_{0}
)f$, where $A_{0}=
\dot\theta$ is an emergent gauge field
coupled to the number operator of the f-electrons.
The path-integral approaches
suggest that the right action for the f-electrons contains an
additional topological term that controls the irreducible 
representation of the spin of the form 
${\cal  L}_{f}= f\dg i\partial_{t}f + A_{0} (n_{f}-Q)$ where $Q=1$ 
for the $SU (2)$ Kondo model.  With this formulation we can
always choose a gauge where $V (t)$ is real and the phase
fluctuations are entirely absorbed into $A_{0} (t)$. The 
NRG results suggest 
that the mean-field saddle point describing a fractionalized ground-state
in which $A_{0}=0$ captures the essential physics of the excitations 
of the  $S=1/2$ SU(2) Kondo model. Fourier transforming the 
conduction electron self-energy into the time-domain,
we see it exhibits long-range temporal correlations 
\begin{equation}\label{}
\Sigma (t_{1}-t_{2})\xrightarrow{|t_{1}-t_{2}|\rightarrow \infty } 
|V|^{2}\hbox{sgn} (t_{1}-t_{2})/2.
\end{equation}
In the single-channel Kondo model
these correlations do not break any physical symmetry but they will be
important for our subsequent discussion. 

\noindent {\bf Induced Order Fractionalization.}
To address our original question regarding order fractionalization, we next
turn to the two-channel Kondo (2CK)
model where the channels of screening 
electrons are indexed by $\lambda=1,2$.
The channel-symmetric two channel Kondo model has a quantum critical
ground-state \cite{Natan,Tsvelik1984,Affleck}
, {which can be loosely interpreted as a resonant-valence-bond (RVB) state of singlet between the spin and each of the two conduction channels (Fig.\,\ref{fig3}{\bf A}). } As we now show, 
breaking this channel symmetry induces
order fractionalization. 
The 2CK exchange interaction is
\begin{equation}\label{}
H_{I}=J\psi\dg_{\lambda}(\vec\sigma\cdot\vec S)\psi_{\lambda} + \delta J\hat {\cal O},
\end{equation}
where the channel asymmetry $\delta J$ couples
to the composite operator
$\hat {\cal O}  = 
(\psi \dg_{1} \vec{ \sigma }\psi _1
- 
\psi \dg_{2} \vec{ \sigma }\psi _2)\cdot\vec S
$, inducing an asymmetric Kondo coupling $J\pm \delta J$ in the
two channels. $\delta J$ plays the role of an external field that induces 
composite order $\langle {\cal O}  \rangle $.
Renormalization group studies tell us that a finite channel asymmetry
$\delta J>0$ destabilizes the quantum critical point, stabilizing
a Kondo singlet in the strongest channel
 \cite{Nozieres80}, here associated with the $\lambda=1$, {loosely interpreted as a valence bond solid (VBS) state} 
(Fig. \ref{fig3}{\bf B}). 
As in the one-channel Kondo model, this implies the formation of a 
fermionic bound-state
\begin{equation}\label{5}
J(\vec{\sigma }_{\alpha \beta }\cdot
\contraction{}
{\vec{S}}
{)}
{\psi }
\vec{S})
\psi_{\lambda \beta } = \bar V_{\lambda} \hat f_{\alpha }, \qquad (\lambda=1,2),
\end{equation}
but with a channel-dependent
amplitude $\bar V_{\lambda}= (\bar  V,0)$ that  projects into the strongest
channel.  The quantum numbers of the
composite fermion divide into two: a 
c-number spinor $\bar V_{\lambda}$ that carries the channel
quantum number and a residual fermion with
spin and charge but no channel index. 

\fg{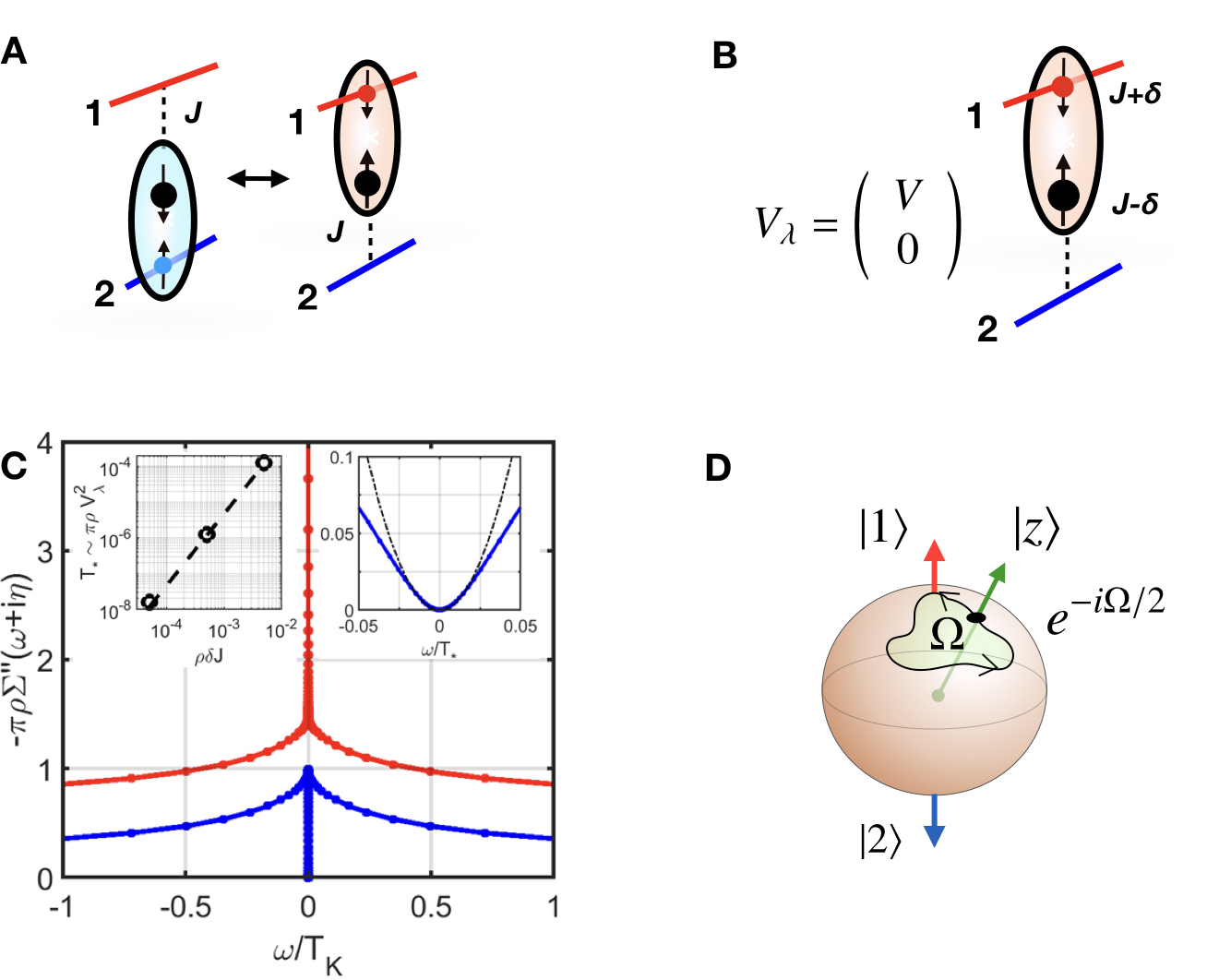}{fig3}{{\bf (A)} The symmetric two-channel Kondo model forms
a quantum critical state where the Kondo singlets are delocalized between channels. {\bf (B)} Application of channel asymmetry stabilizes
a Kondo singlet on one channel, defined by a spinor order parameter. 
{\bf (C)} 
A fermionic pole is induced in
the  strongest channel, indicating the presence of a selective
hybridization $\bar V_{\lambda}= (V,0)$ between the fractionalized moment
and the two screening channels. 
The red curve is shifted by 0.5 unit vertically upward for clarity. (Insets) left: $T_*$ vs $\rho\delta J$. right: the low-frequency
regime of self-energy in channel 2 vs. $\omega$ showing Fermi liquid
behavior (dashed black line is an $\omega^2$ fit). {\bf (D)} Adiabatic rotation of the channel asymmetry 
along a path subtending a solid angle $\Omega$ leads 
between channels leads to an $e^{-i\Omega/2}$ Berry phase, reflecting the
half-integer nature of the induced order.
 }

Fig. \ref{fig3}{\bf C} displays the self-energy
of a channel-asymmetric spin-$1/2$ 2CK model, calculated using
NRG. For $\delta J>0$, 
a sharp, resolution-limited 
quasiparticle pole forms in the strongest $\lambda=1$ channel, leading to a pole in the
self-energy, 
 $\Sigma_{\lambda\lambda'}\sim
\bar V_{\lambda}V_{\lambda'}/\omega$. The product form of the
self-energy follows from the projection into the strongest channel. 
By Fourier transforming this result, we confirm that at long times, 
the self-energy factorizes 
\begin{equation}\label{}
\Sigma_{\lambda\lambda'} (t_{2}-t_{1}) 
\xrightarrow{|t_{2}-t_{1}|
\rightarrow \infty }
\bar V_{\lambda} (t_{2}) V_{\lambda'} (t_{1}) 
 {\rm
sgn } (t_{2}-t_{1})/2.
\end{equation}
This factorization into two spinors 
means that the singular part of the self-energy does not transform under an irreducible
representation, but instead as a {\it reducible} sum of
a vector and scalar representation, i.e 
$1/2\otimes 1/2 = 0 +1 $. If we are to preserve Landau's notion that
order parameters transform under irreducible representations, then we
are forced to acknowledge that the spinor $V_{\lambda}$ is the
relevant order parameter and we have order fractionalization.

A way of exposing the spinorial character of the order is to
adiabatically rotate the spinor $V_{\lambda}$
by slowly rotating the channel asymmetry, which
we may write $\delta J (\hat n (t)\cdot \vec{{\cal O} })$, where
$\vec{{\cal O}} = \psi \dg_{\lambda} \vec{\alpha
}_{\lambda\lambda'} (\vec{\sigma} \cdot \vec{S})\psi_{\lambda'}$ is
the composite ``channel magnetization'', defined in terms of three Pauli matrices
$\vec{\alpha }= (\alpha_{1},\alpha_{2},\alpha_{3})$, 
and $\hat n (t)$ is the asymmetry field. 
In this adiabatic evolution of the Kondo singlet, 
the channel selective hybridization is then 
determined by  $(\vec{\alpha }_{\lambda\lambda'}\cdot
\hat n )V_{\lambda'} = +V_{\lambda}$. 
By slowly varying $\hat n (t)$ along a closed path, we can characterize the topology of the order parameter by examining the corresponding
Berry phase factor $e^{-i\gamma }
=e^{-i\Omega S}
$ where $\Omega$ is the solid angle enclosed by the path and $S$
determines the spin of the  order parameter (Fig. \ref{fig3}{\bf D}). 
To calculate $\gamma $ we go to 
the extreme strong coupling limit, where $\delta
J\gg\Lambda$, the electron band-width. In this limit the channel Kondo singlet
becomes entirely local, taking the form 
$\ket{z}=\sum_{\lambda =1,2}z_{\lambda} \vert \lambda_{s}\rangle $, 
where $z= (z_{1},z_{2})\equiv V_{\lambda}/|V|$ is a unit spinor and $\vert \lambda_{s}\rangle  =
\frac{1}{\sqrt{2}}\left(\vert \lambda\uparrow,\Downarrow \rangle - \vert
\lambda\downarrow , \Uparrow\rangle  \right)
$ denotes a singlet formed between the local moment ($\Uparrow$) 
and electron ($\vert \lambda\uparrow\rangle $) in channel $\lambda$.
When $\hat n (t)$ is rotated through a solid angle $\Omega$, the
spinor $z_{\lambda}(t)$ evolves adiabatically, and the ground-state wavefunction
acquires a Berry phase given by $\gamma = -i\oint dt ( z\dg \partial_{t}z) =
\frac{1}{2}\Omega$ (cf. Supplementary Material C). 
The factor of $1/2$ confirms the half-integer
channel spin of the state. 

We also note the
application of the symmetry-breaking field $\delta J$ induces an
expectation value of the composite order parameter
$\langle  \vec{{\cal O} }\rangle 
\propto V^{*}_{\lambda}\vec{\alpha }_{\lambda\lambda'}V_{\lambda'}$,
indicating  that the induced composite order has fractionalized into a
product of channel hybridization spinors.
From this exercise, we see that order fractionalization (OF) is manifested
in three separate ways:  (i) a splitting of the
composite fermion into a spinor boson and a
fermion; (ii) a fractionalization of the local
moment into a product of fermion excitations and (iii) 
fractionalization of the composite order parameter into a product of spinorial
order parameters. 




\figwidth=\textwidth
\fg{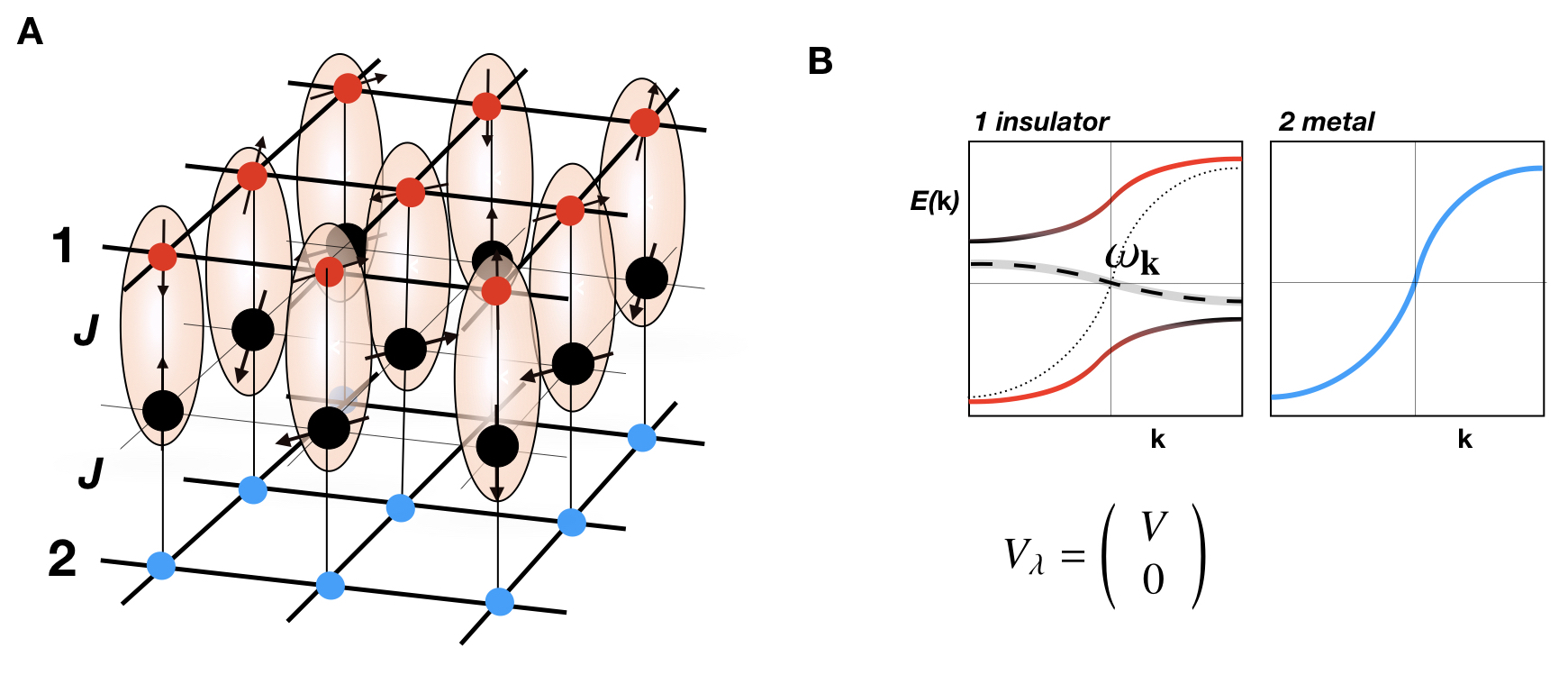}{Fig4}{{(\bf A)} In a 2CK lattice model at $d=2,3$
dimensions we expect order fractionalization to lead to
spontaneous channel symmetry breaking, forming Kondo singlets in one
channel, with  a channel-spinor defining the 
selective hybridization of the fractionalized local
moments. {(\bf B)} The channel spinor leads to a Kondo insulator in
the hybridized channel (1) with a hybridization gap.  The dashed line
indicates the dispersion $\omega_{\bk }$ of the emergent
f-electron. Channel 2 remains a gapless insulator. 
}

%

\noindent {\bf Spontaneous Order Fractionalization.} Induced order fractionalization in the two-channel Kondo impurity 
model enables us to 
conjecture  spontaneous order fractionalization in lattice models.  
The key idea is that order fractionalization at one site produces a
channel-symmetry breaking Weiss field at its neighbors. This then provides 
positive feedback that allows a fractionalized ordered state to
develop spontaneously due to the fragile nature of the critical system.
The NRG results confirm that the path integral gauge theory
approach  provides a qualitatively correct description of the 
fractionalization at a mean-field level, 
which can then be used to describe spontaneous
order fractionalization in a Kondo lattice. As in conventional
mean-field theories, the Gaussian order parameter 
fluctuations about such a mean-field theory are finite in 
dimensions $d\ge 2$, allowing a stable
spontaneous order fractionalization 
 \cite{Wugalter18,Zhang18}. Thus in the two-channel Kondo lattice at
half filling, spontaneous OF means that the system behaves as a 
Kondo insulator in one channel, remaining metallic in the other. 
With this
reasoning, symmetry-breaking
phases of this type characterized in dynamical mean-field
theory calculations \cite{Jarrell96,Nourafkan08,Hoshino11} can be 
interpreted as order-fractionalized phases. 


\noindent {\bf Patterns of Fractionalization}. In our 
studies we have highlighted just one example of how broken channel
symmetry induces an order fractionalization. 
By varying the
asymmetry field ${\cal O}$ we can classify family of fractionalized
order, which we have summarized in Table \ref{table1}.  Class (a) is
fractionalization of the single channel Kondo model, with no broken
symmetry. Class (a') is discussed below. 
Our NRG calculation
gives us a channel FM, listed as class (b). 
To obtain the other patterns of fractionalization, we 
we have emplyed two different strategies.
The first is to take advantage of the high symmetry of the
2CK to rotate the pattern of fractionalization. 
In the 2CK the two channel symmetry and particle-hole symmetries form
part of single SO(5) symmetry \cite{Affleck92,zhangso5,Kuramoto} (see
supplementary material E).  This allows us to rotate the asymmetry field
$\vec{{\cal O}}= ({\cal O}_{1},\dots {\cal O}_{5}) $ within a five dimensional
space, in which the first three components correspond to channel magnetization
as discussed above, while the fourth and fifth  correspond to
a composite pair operator ${\cal O}_{4}-i {\cal O}_{5} \equiv
\psi_{1}\vec{\sigma }i\sigma_{2}\psi_{2}\cdot \vec{S}$. 
When we rotate the asymmetry field 
from a channel magnet into a composite pair, we can follow the
corresponding transformation in the composite fermion, showing that it
acquires a Boguilubov structure as shown in class (c).
Finally, we can obtain a fifth class (d) by reversing the roles 
the roles of channel and spin to obtain a quadrupolar Kondo model.  Here 
the fractionalized excitations carry the channel index $\bigl
[(\vec S\cdot\vec\sigma)\psi\bigr ]_{\lambda\alpha} \to V_\alpha
f_\lambda$ while the order parameter $V_\alpha$ is a spinor
that breaks time reversal symmetry. 

Each of the examples of order fractionalizalization 
apply to two-channel Kondo
models. We can employ a second strategy to examine order
fractionalization in one channel Kondo systems, by
using a reformulation of the 
2CK in terms of Majorana fermions
in which the spin-density of the two channels
is replaced by the spin and isospin density of a ``compactified''
two-channel Kondo
model, 
$
(\psi \dg_{1}\vec{\sigma }\psi_{1}, \psi \dg_{2}\vec{\sigma }\psi_{2})
\rightarrow 
(\psi \dg\vec{\sigma }\psi, \tilde{\psi }\dg\vec{\tau }\tilde{\psi})
$ \cite{colemanioffe} where $\tpsi$ is a Nambu spinor. In this form, the fractionalization can analyzed
(see supplementary materials D) by 
studying the family of Kondo interactions $J\tpsi (\vec{\sigma
}+ (1-\epsilon)\vec{\tau })\tpsi \cdot \vec{S}$, where $\epsilon\neq 0$
breaks the channel symmetry.  
This formulation can be examined in the
strong coupling limit, demonstrating that the local moment
fractionalizes into Majorana, rather than Dirac 
fermions \cite{Coleman94,Xu10}, as shown in
entry a' of Table 1. 

\input{table1Q.tex}

Candidate materials with 
composite order corresponding to these different fractionalization
patterns have been proposed in the context of heavy fermion materials.
materials. 
The SO(5) rotations of the channel FM into a composite ordered state
\cite{Coleman:1999vb} (Table 1, class c) has been proposed for 
the heavy fermion superconductor
NpPd$_{2}$Al$_{5}$ \cite{Flint:2008fk}.  
An example of class (d) is the ``hastatic''
hidden order proposed for phases of URu$_{2}$Si$_{2}$
 \cite{Chandra2013} and PrTi$_2$Al$_{20}$  \cite{Dyke18}. 
The Majorana fractionalization class (a')
has been recently proposed
as a candidate for the strange insulating state in SmB$_{6}$
\cite{Erten:2017b}. 
Each of these  materials are candidates for 
the realization of spontaneous order fractionalization. 

\noindent {\bf Order Fractionalization Conjecture.}
We can also envisage spontaneous 
order fractionalization in broader contexts beyond 
Kondo lattices; for example in the Hubbard model where 
the combination of spin and
fermion is now replaced by a three fermion bound-state.   
This leads us to conjecture that in this more general situation, 
the OF will involve
a fractionalization of a three-fermion bound-state
into two components:
a bosonic ``{\it corona }'' surrounding
a {\it ``dark  fermion''} located at the center-of-mass.
This process has
the effect of partitioning the 
quantum numbers $\Lambda= (\{\lambda\} ,\{\alpha\} )$  of three-body composite, 
into two parts, 
the $\lambda$ variables
reside exclusively in the bosonic corona, while the $\alpha $ variables
reside in the dark fermion. The order fractionalization conjectured
(OFC) 
for this
general case takes the form
\begin{equation}\label{}
\bigl(\mathrel{\mathop
{{
\psi
\psi
\psi 
}}
^{\dsp \ltdash
\joinrel\relbd
\joinrel
 \ltdash
\joinrel\relbd\joinrel
 \ltdash
}}\bigr )
_{\Lambda} (x) = V^{\lambda}_{\alpha \alpha'} (x) f_{\alpha'}(x).
\end{equation}
Here $(\psi \psi\psi )_{\Lambda} (x)$ corresponds to a combination of
creation or annihilation operators with center of
mass $x$, that transform under 
fundamental representations of the $\Lambda$. $V^{\lambda}_{\alpha
\alpha '} (x)$ and $f_{\alpha } (x)$ are the order parameter corona
and the dark fermion respectively. In the simplest cases,
$V^{\lambda}_{\alpha \beta }= V^{\lambda}\delta_{\alpha \beta }$ is
diagonal, and the corona and dark fermion share a common $U (1)$ gauge
symmetry.  The general matrix structure of the order fractionalization
allows for a non-abelian partition of the quantum numbers, with an internal
SU(N) gauge symmetry associated with the quantum
numbers $\alpha $.  This more general form is required to understand
the example of composite pairing shown in Table. \ref{table1}.

The OFC also  implies that the corresponding self-energy
factorizes as follows
\begin{equation}\label{}
\Sigma_{\Lambda \Lambda'} (2,1) \xrightarrow{|2-1|
\rightarrow \infty }
\bar V_{\lambda} (2) g (2-1)V_{\lambda'} (1).
\end{equation}
where $g(2-1)$ is the one-particle propagator of
the dark fermion. 
In a lattice, the dark fermions will generically delocalize with
dispersion $\omega_{\bk }$, forming a 
Fermi surface  ${\bf k }\in \{\bk_{F}^{*} \}$ where $\omega_{\bk_{F}^{*}}=0$ vanishes.
In space time, the 
asymptotic Green's function of the dark Fermions, 
\begin{equation}\label{}
g(\vec{x},t)\sim \delta_{\alpha\alpha '}
\frac{e^{i \vec{k}_F^{*}  \cdot \vec{x}}}{x - v_F(\hat x)t },
\end{equation}
where $\vec k_F^{*} (\hat x)= {\bk  }_{F}^{*}$ is the Fermi wavevector at the extremal
point on the Fermi surface where the group velocity 
$\vec{ v}_{F}= v_{F}\hat x$ is parallel to $\vec x$. 
This defines a kind of ``light cone'' on which $g$ is arbitrarily large.
The
factorization of the self-energy 
into a product of spinors is the conjectured
outcome of order fractionalization in a fermionic system, and
constitutes a generalization of the concept of off-diagonal long range
order into the time domain. 
We also note that the singularity  $\Sigma_{\lambda,\lambda'} (\omega, \bk_{F}^{*}) \sim
\bar  V_{\lambda}V_{\lambda'}/(\omega-\omega_{\bk } )$
in the self-energy at the dark Fermi
surface leading to zeroes in the electronic Green's function
$G(\omega,\bk_{F}^{*})=0$  \cite{Phillips7,seki17}.  There is an
interesting possible link  with singularities in the electron self-energy
observed in cluster dynamical mean-field studies of the Hubbard model  \cite{Sakai16} 
and also proposed as a phenomenological explanation of Fermi arcs in
under-doped cuprate superconductors  \cite{Rice06,Konik:2006cb}. 


Conventional and fractionalized
order can be delineated in various ways. There are a number of quantum materials, including NpPd$_{5}$Al$_{2}$,
CeCoIn$_{5}$, UBe$_{13}$, PrV$_{2}$Al$_{20}$
and PrTi$_{2}$Al$_{20}$ where spin or quadrupolar Kondo effects 
coincide with phase transitions into broken symmetry states. 
An important ``fingerprint'' of fractionalization is the appearance 
of dark fermionic bound-states, 
that may be detected using 
spectroscopies such as angle resolved photoemission (ARPES) or
scanning tunneling microscopy (STM).  For example if in CeCoIn$_{5}$ the
Kondo effect coincides with the development of superconductivity, 
then STM should detect an expansion of the Fermi surface at the 
superconducting transition; in neutral cases the thermal conductivity
would be an ideal probe for this Fermi surface change.

An intriguing question is whether the different topologies of 
fractionalized order can be detected
experimentally.  For example in  UBe$_{13}$ and  Pr (V,Ti)$_{2}$Al$_{20}$ 
where the channel index is spin, it may be possible to 
externally manipulate the 
channel-symmetry breaking Kondo-effect: rotating the order spatially 
through 360$^{\circ }$ to create a $\pi$ phase shift may be
detected in a channel interferometer; rotating the order in time using
optical methods may lead to a breathing Fermi surface which might be
measured using a channel-selective conductivity.

We end by noting that despite the spin-statistics theorem,
relativistic versions of order fractionalization are possible since 
Lorentz invariance does not prohibit bosons that carry 
half-integer isospin.  A classic example is the Higgs boson, 
a spinor which carries half-integer weak isospin and could
conceivably emerge as a fractionalized order parameter of more
fundamental Fermi fields. 

In conclusion we have presented a mechanism for the fractionalization of
order parameters through the formation of fermionic poles in the
self-energy; this enables long-time factorization of the self-energy
into products of order parameters transforming under the fundamental
representation of the symmetry group. We have provided crucial substantiating
examples in the single and two-channel impurity Kondo models.  These
results have led us to conjecture that such
phenomena may appear spontaneously in a lattice as suggested by
several experimental, mean-field and computational results. 

We would like to thank 
C. Batista, M. Civelli, R. Flint, E. K\"onig, M. Imada, M. Oshikawa, P. Phillips, A. Rosch, S. Thomas, S. Sondhi,
S. Sachdev, T. Senthil and A. Wugalter  for stimulating discussions. The
conceptual beginnings of this work were 
supported by NSF grant DMR-1309929 (P. Coleman). 
Yashar Komijani is supported by a Center for Material Theory
postdoctoral Fellowship. 
We thank Princeton
University, Princeton USA (P. Chandra) the Flatiron
Institute, New York USA  (Y. Komijani and P. Coleman) 
and the Institute for Solid State Physics, Kashiwanoha, Japan 
(P. Chandra and P. Coleman) for their
hospitality during the later writing stages of this manuscript.
\input{suppl.tex}

\bibliography{ofcx}

\end{document}

%% file: header.tex
\usepackage{xcolor}
\usepackage{amsmath,amssymb,bbold,bm}
\usepackage{graphicx}
\usepackage{mathdots}
\usepackage{caption}
\usepackage{subcaption}
\usepackage{cancel}
\usepackage{comment}
\usepackage{currfile}

\newcommand{\be}{\begin{equation}}
\newcommand{\ben}{\begin{equation*}}
\newcommand{\ee}{\end{equation}}
\newcommand{\een}{\end{equation*}}
\newcommand{\bs}{\begin{split}}
\newcommand{\es}{\end{split}}
\newcommand{\bmx}{\begin{array}}
\newcommand{\emx}{\end{array}}
\newcommand{\bea}{\begin{eqnarray}}
\newcommand{\bean}{\begin{eqnarray*}}
\newcommand{\eea}{\end{eqnarray}}
\newcommand{\eean}{\end{eqnarray*}}
\newcommand{\dg}{^{\dagger}}
\newcommand{\dn}{^{\vphantom{\dagger}}}

\newcommand{\ua}{\uparrow}
\newcommand{\da}{\downarrow}

\newcommand{\qqquad}{\qquad\qquad\qquad}
\newcommand{\so}{\qquad\rightarrow\qquad}

\newcommand{\eps}{\epsilon}

\newcommand{\tpsi}{\tilde{\psi}}

\newcommand{\sgn}[1]{{\rm sign}{#1}}
\newcommand{\pref}[1]{(\ref{#1})}

\newcommand{\tr}[1]{{\rm Tr}\Big[ #1 \Big]}

\newcommand{\abs}[1]{\left\vert #1 \right\vert}
\newcommand{\bra}[1]{\left\langle #1 \right\vert}
\newcommand{\ket}[1]{\left\vert #1\right\rangle}

\newcommand{\braket}[1]{\left\langle #1\right\rangle}

\newcommand{\mat}[1]{\left(\bmx{cc}#1\emx\right)}
\newcommand{\matc}[2]{\left(\bmx{#1}#2\emx\right)}

%% file: table1Q.tex
\vskip 0.1in
\begin{table}[ht]
\begin{tabular}
{|l||c|c|c|c|c|c|}
\hline
Kondo Model&3 body state&Class&Composite Fermion& Induced Order & Asymmetry ${\cal O}$ &OF\\
\hline
\multirow{4}{*}{One Channel}
&\multirow{4}{*}{$(\vec{\sigma }\cdot
\contraction{}
{\vec{S}}
{)_{\alpha \beta }}
{\psi }
\vec{S})_{\alpha \beta }
\psi_{\beta }
$
}
&\multirow{2}{*}{a}
&\multirow{2}{*}{$ {V}\ f_{\beta }$}
&\multirow{2}{*}{(Fermi Liquid)}
&\multirow{2}{*}{ \textemdash}
&\multirow{2}{*}{ \textemdash}
\\
&&&&&&
\\
\cline{3-7}
&&\multirow{2}{*}{a'}&
\multirow{2}{*}{$(\vec \sigma \cdot \vec{ \eta })_{\alpha \beta }{\cal V}_{\beta
}$}
&{Odd $\omega$ pairing}
&
\multirow{2}{*}{$\psi_{\uparrow}\psi_{\dw}S^{+}+{\rm H.c}$}&\multirow{2}{*}
{${\cal V}_{\up}{\cal V}_{\dw}+ {\rm H.c}$ }
\\
&&&\cite{balatsky95,Coleman94}&&&\\
\hline
\multirow{8}{*}{Two channel}&Spin&
\multirow{2}{*}{b}&
\multirow{2}{*}{$V_{\lambda }f_{\alpha }$
}
&{Channel FM}&
\multirow{2}{*}{$\psi_{\lambda} \dg (\vec{\sigma }\cdot
\vec{S})\psi_{\lambda'} $
}
&
\multirow{2}{*}{$\bar {V}_{\lambda} V_{\lambda'}$
}
\\
&\multirow{3}{*}{$(\vec{\sigma }\cdot
\contraction{}
{\vec{S}}
{)_{\alpha \beta }}
{\psi }
\vec{S})_{\alpha \beta }
\psi_{\lambda\beta }
$
}
&&\cite{Hoshino11,Dyke18}&&&\\
\cline{3-7}
&&
\multirow{2}{*}{c}&
\multirow{2}{*}{$V_{\lambda}f_{\alpha }+ \Delta_{\lambda}\tilde{\alpha }f\dg
_{-\alpha }$
}&Composite pairing&
\multirow{2}{*}{ $\psi_{1}(\vec{\sigma }\cdot \vec S)\sigma_{2}\psi_{2}
$
}
&
\multirow{2}{*}{ $
V_{1}\Delta_{2}-V_{2}\Delta _{1}$
}
\\
&&
&&\cite{emery92,Coleman:1999vb}&&
\\
\cline{2-7}
& Quadrupolar&&&&&\\
&
\multirow{2}{*}{$(\vec{\gamma}\cdot
\contraction{}
{\vec{S}}
{)_{\lambda\lambda'}}
{\psi }
\vec{S})_{\lambda\lambda' }
\psi_{\lambda'\alpha }
$}&
d
&$V_{\alpha }f_{\lambda}$
&{Hastatic Order}&
$\psi_{\alpha } \dg \vec{\sigma } (\vec{\gamma}\cdot\vec {S})
\psi_{\beta } 
$
&
$\bar {V}_{\alpha }V_{\beta }
$
\\
&&&&\cite{Chandra2013}&&\\
\hline
\hline
\end{tabular}
\caption{\label{table1} {\small 
Classifying different patterns of fractionalization in single-channel and
two-channel Kondo systems, obtained by adding asymmetry fields to one
and two-channel Kondo models. 
In class a'  $\vec{\eta } = (\eta_{1},\eta_{2},\eta_{3})$ denotes a vector of three
emergent Majorana fermions that fractionalize the local moment spin,
$\vec{S}\equiv \frac{i}{2}\vec{\eta }\times \vec{\eta
})$\ \cite{Coleman94}. Class b channel rotations
of the asymmetry field $\cal O$, whereas  class c are obtained from an
SO(5) rotation of the channel asymmetry (Supplementary Material XXX). 
In class c $\tilde{\alpha }$ denotes  $\tilde{\alpha }\equiv \sgn{\alpha}$. 
}}
\end{table}

%% file: suppl.tex
\newcommand{\nocontentsline}[3]{}
\newcommand{\tocless}[3]{\bgroup\let\addcontentsline=\nocontentsline#1#2{#3}\egroup}
\tocless\section*{Supplementary Material} 
This supplementary material contains
additional details and proofs for key statements in the paper. Section A contains details of the Numerical Renormalization Group (NRG) calculation.  Section B
contains a derivation of the fermionic pole in the self-energy of the
Kondo problem using Fermi-liquid theory, 
in the single-channel and channel-asymmetric two-channel Kondo
models. Section C contains a proof of the Berry phase accumulated by the ground state under an adiabatic time-evolution, establishing the spinorial character of the order parameter. Section D shows that the projective form of the self-energy and the fractionalization pattern can be directly extracted by studying the the strong coupling limit of single-channel Kondo problem for both usual and the compactified model. In section D we review the SO(5)$\times$SU(2) symmetry group of the two-channel Kondo problem and use the extended symmetry to study various patterns of fractionalization.\\

\tableofcontents

\subsection{Details of the NRG calculations} NRG calculations were 
performed using the density-matrix NRG code \cite{Toth08,Toth08b} with a flat
density of states, which produces the imaginary part of the local
Green's functions (e.g. $G(z)=\langle\langle
\psi_{\sigma};\psi_{\sigma}\dg\rangle\rangle_z$) at discrete frequencies
determined by the Wilson discretization as follows 
\be
-G''(\omega+i\eta)=\frac{1}{Z}\sum_{n,m}\vert
\braket{m\vert \psi_{\sigma}\dg \vert n}\vert^2
\pi\delta(\omega+E_n-E_m) (e^{-\beta E_{m}}+e^{-\beta E_{n}})
,\label{eq1} \ee 
where $\ket{m}$ and $\ket{n}$ are many-body eigenstates with energies $E_{m}$ and $E_{n}$, respectively, and $Z= \sum_n
e^{-\beta  E_{n}}$ is the partition function. 
Our methodology takes account of the 
SU$_{charge}$(2)$\times$SU$_{spin}$(2) 
and SU$_{spin}$(2)$\times$SU$_{charge
1}$(2)$\times$SU$_{charge 2}$(2) symmetries of the one and the
two-channel models, respectively to simplify calculation of the matrix
elements. 
The calculations employed 500 multiplets and
a Wilson parameter $\Lambda=1.8$, with a Wilson chain length $L=80$. 
The full-density matrix
calculation ensures that the discrete results \pref{eq1} satisfy the
sum-rules enforced by the commutation relations. The code uses
an interpolative log-Gaussian broadening, with the Kernel defined as follows
\cite{Bulla01,Toth08} 
\begin{eqnarray}\label{l}
K_{int}(\omega,\omega_i)&=&\frac{1}{b\sqrt{\pi}}e^{-[x(\omega)-x(\omega_i)]^2/b^2}\frac{dx}{d\omega_i},\cr
x(\omega)&=&\frac{1}{2}\tanh({\omega}/{T_Q})\log[(\omega/T_Q)^2+e^{\eta}]
\end{eqnarray}
where the `quantum temperature' $T_Q$ is chosen to be $
10^{-15}$ and $\eta\approx T_Q$ for our zero temperature
calculations. We have taken the logarithmic broadening 
parameter $b$ to be $b=0.6$.
This leads to 
\be 
-G''(\omega+i\eta)=\frac{1}{Z}\sum_{n,m}\vert
\braket{m\vert \psi_{\sigma}\dg \vert n}\vert^2
\pi K(\omega, E_m-E_n) (e^{-\beta E_{m}}+e^{-\beta E_{n}}).  \ee 
The normalization 
\be
\int{d\omega}K_{int}(\omega,\omega_i)=1 \ee 
ensures that the sum-rules
are satisfied.  Following the standard approach, the Hilbert
transform is applied to the broadened data to obtain both real and
imaginary parts of the Green's function.

We compute the irreducible self-energy for
the Kondo problem from 
the conduction electron Green's function $G
(z)$. In principle, the self-energy can be calculated directly from
the relation
\begin{equation}\label{}
\Sigma (z)= g^{-1} (z)- G^{-1} (z)
\end{equation}
where $g (z)$ is the bare conduction electron propagator in the
absence of the impurity ($J=0$). However, at the Fermi energy $G^{-1} (z)$ is
singular and so this expression requires careful regularization.

In practice, we found it easier to divide the calculation into two
parts. First
we calculated the electron T-matrix $T (z)$, defined by the
relation
\be
G(z)
=g(z)+g(z)T(z)g(z)
\ee
from which we obtain
\begin{equation}\label{}
T (z) = \frac{G (z)-g (z)}{g (z)^{2}}.
\end{equation}
The regularized self- energy  was then calculated from the T-matrix,
using the relationship
\begin{equation}\label{}
\Sigma(z)=\lim_{\eps\to
0}\hphantom{A}\frac{T(z)}{1+(1-\eps)g(z)T(z)},\label{eq11} \qquad
g(z)=\langle\langle \psi\dn_\sigma;\psi\dg_\sigma
\rangle\rangle_z\Big\vert_{J=0}.
\end{equation}
This limiting procedure was used to
ensure that the $\Sigma(z)$ has the correct analytical properties. 
A
finite $\eps$ leads to some residual broadening of the peak. In our
calculations, we used $\eps=10^{-8}$.

\subsection{Fermionic pole in the Kondo self energy: relationship to Fermi-liquid theory} 
\begin{figure}[h!]
\includegraphics[width=0.8\linewidth]{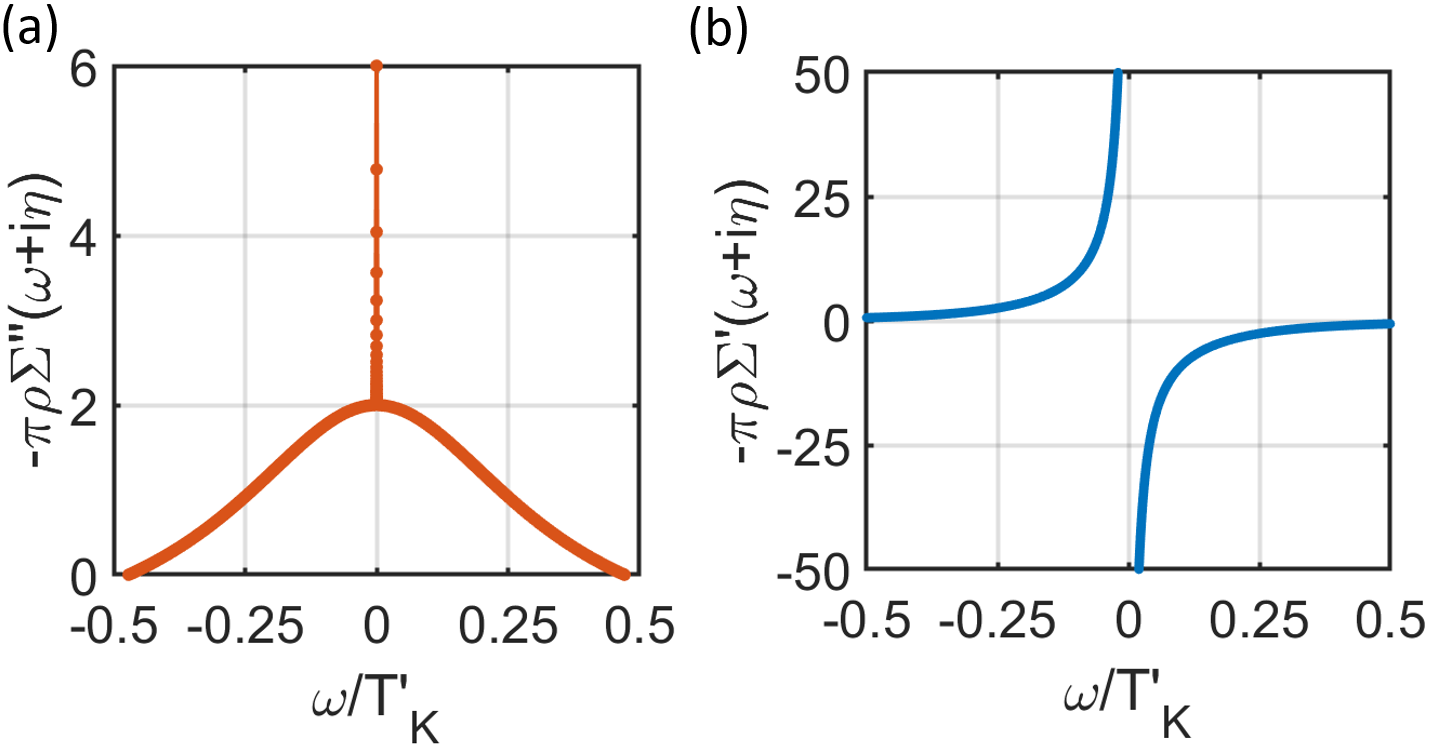}\caption{\small (a) The imaginary part of the 
irreducible self-energy extracted from Fermi-liquid theory. The
background is a result of the inelastic scattering of the composite
quasiparticle at finite energy as shown in Eq. (20). (b) The real part shows the $1/\omega$ signature of long-range order in time.}\label{figFL}
\end{figure}

We can gain
analytic insight into our results using Fermi-liquid theory. 
From Fermi-liquid
theory \cite{Affleck,Shi} we know that the phase shifted
quasi-particles have a Fermi-liquid interaction with a single scale: the Kondo temperature.
This gives rise to a 
low energy scattering $T$-matrix of the form
\begin{equation}
-\pi\rho
T(\omega+i\eta)=i-\frac{\omega}{T'_K}-i\zeta\frac{\omega^2}{T_K'^2} 
+O (\omega^{3}), 
\label{eq10}
\end{equation}
where $T'_K$ is proportional to the weak-coupling
Kondo temperature $T_K\sim D\sqrt{J\rho }e^{-1/\rho J}$. 
Inserting  this expression, together with 
a flat density of states $g(\omega+i\eta)=-i\pi\rho$ into (18), the
corresponding irreducible self-energy is then
\begin{equation}\label{}
\pi\rho\Sigma(\omega+i\eta)= \Big[{\frac{1}{\pi\rho T (\omega+i\eta )}-i}\Big]^{-1}=
\frac{T'_K}{\omega
+i(\zeta-1)\omega^2/T'_K +i\eta}.
\end{equation}
This result is plotted in Fig.\,\ref{figFL}.  The background beneath the delta-function pole
is thus understood as a result of the $\omega^{2}$ Fermi liquid
scattering rate. 
The parameter $\zeta$
determines the value of background offset $-\pi\Sigma''(\omega\to
0)=\zeta-1$, and does not affect any
qualitative features of the discussion. In order to recover our
numerical results $-\pi\Sigma''(\omega\to 0)=2$, one must set $\zeta=3$
here. This is a factor of two larger than the value of $\zeta$
derived in \cite{Affleck}. The origin of this discrepancy is unclear
to us at this point. 

For the two-channel Kondo impurity model with channel symmetry, the
T-matrix at $T=0$ is given by \cite{Affleck} \be -\pi\rho
T(\omega+i\eta)=i/2+O (\sqrt{\omega}) \ee which is responsible for the
$-\pi\rho\Sigma''(0+i\eta)=1$ at the 2CK fixed point. In the presence
of channel asymmetry, the system flows to a local Fermi liquid fixed
point with resonant scattering in the strongest channel.  The Fermi
liquid temperature is given by \cite{Pustilnik04} \be T_*=T_K\kappa^2,
\qquad \kappa^2=4\frac{(\rho J_1-\rho J_2)^2}{(\rho J_1+\rho J_2)^4},
\ee in the scaling limit $D\to\infty$ and $\rho J_i\to 0$. The latter
limit has to be taken such that $\kappa^2$ remains finite, so that
$T_*/T_K$ remains finite.  Our extracted $T_*$ vs. $\rho\delta J$ is
shown in the left inset of Fig.\,3{\bf B} and it agrees with this
formula (shown as a broken line). At $T=0$ and $\abs{\omega}\ll T_*$
the self-energy in the two-channels are distinctively different. The
self-energy of the stronger channel contains a sharp pole on top of
the Fermi liquid background, whereas the weaker channel only contains
a Fermi-liquid $\omega^2$ contribution. The result can be fit with the
following formula:
\be
-\pi\rho\Sigma(\omega+i\eta)=\mat{-[(\omega
+i\eta )/T^{*} +2i\omega^2/T_*^2]^{-1}
& 0 \\ 0 & 2i\omega^2/T_*^2}, \qquad \ee

\subsection{Berry phase calculation}

In this section, we calculate the Berry phase associated with a slow time-dependent
change in the channel asymmetry of the two channel Kondo model.  Since
the Berry phase is a topological quantity, it is independent of
coupling strength, and we 
can carry out this calculation in the strong-coupling limit of the model,
given by
\[
H[\hat n]= H_{I}+ \Delta J \hat n\cdot\vec{{\cal O}}
\]
where $H_{I}= J \sum_{\lambda}\psi \dg_{\lambda}\vec{\sigma
}\psi_{\lambda}\cdot \vec{S}$ is the symmetric Kondo interaction and 
\begin{equation}\label{}
\vec{{\cal O}}=\psi\dg_\lambda \vec\alpha_{\lambda\lambda'}(\vec\sigma\cdot\vec S)\psi\dn_{\lambda'}
\end{equation}
is the ``channel magnetization'', 
where $\vec{\alpha }$ are a set of Pauli
matrices in channel space and 
\be
\vec n(t)=(\sin\theta_t\cos\phi_t,\sin\theta_t\sin\phi_t,\cos\theta_t).
\ee
is the time-dependent asymmetry field. 
We have replaced $\delta
J\rightarrow \Delta J$ to denote a large finite value of the channel
asymmetry. 
Provided $\Delta J$ and $J$ are much
larger than the electron band-width, we can ignore everything except the
one-site Hamiltonian. 
At $\Delta J=0$, the
ground-state is an over-screened local moment with ground-state energy
$E= -2J$. Beyond a critical asymmetry, a  channel asymmetric
singlet state with energy  $-\frac{3}{2} (J+\Delta J)$ is stabilized.  
This requires that $\Delta J> J/3$. 

We can parameterize the asymmetry field  using a CP$^{1}$ representation 
$\hat n= \bar z\vec{\alpha
}z$, where 
\begin{equation}\label{eq14}
z[\hat n]= \mat{z_{1}\cr z_{2}} = \mat{\cos \theta/2\cr\sin\theta/2 e^{i\phi }}.
\end{equation}
Then the 
ground-state for a particular fixed value of $\hat n$ can be
written as 
\begin{eqnarray}\label{l}
\ket{z}&=&\sum_{\lambda=1,2}
\ket{\lambda_{s}}z_\lambda,\cr
\vert \lambda_{s}\rangle  &=&
\frac{1}{\sqrt{2}}\left(\vert \lambda\uparrow,\Downarrow \rangle - \vert
\lambda\downarrow , \Uparrow\rangle  \right)
\end{eqnarray}
Here the $\Uparrow, \Downarrow$ refer to the spin state of the local
moment and $|\lambda \uparrow\rangle, \vert \lambda \downarrow \rangle $ refer to the spin state of the
electron in channel $\lambda$.  At the site of the moment, 
the weaker channel is  either
empty or doubly occupied,  so that its spin is shut down. This charge
state in the weaker channel
$\alpha\ket{0_{\bar{\lambda\dn}}}+\beta\ket{2_{\bar{\lambda\dn}}}$
forms an isospin variable in the charge sector which, since it
commutes with the Hamiltonian, is not relevant for this discussion and
we do not include it in $\ket{\lambda_{s}}$. 

When evolved adiabatically, the spin singlet follows 
the direction of the channel asymmetry.  If the applied field $\hat n
(t=T)=\hat n (t=0)$ returns to its original direction, the state
returns to its original ground-state, up to a finite phase. 
The state at time $t$ can be written as
\be
\ket{\psi(t)}=e^{-i\alpha (t)}\ket{z (t)}
\ee
where 
\be
\alpha(t)=\int_0^t dt' E(t')+\eta(t).
\ee
Here the first term is the Schr\" odinger phase accumulation
associated with the energy, which in our case is constant $E(t')=E$. The second term $\eta(t)$ is the Berry
phase.  Inserting this into the time-dependent Schroedinger equation
\be
i\hbar\partial_t\ket{\psi(t)}=H[\vec n(t)]\ket{\psi(t)}
\ee
we find
\be
\partial_t\ket{z(t)}-i\dot\eta \ket{z(t)}=0
\ee
Multiplying from left by $\bra{\bar z}$ we find
\be
\dot\eta=-i\braket{\bar z\vert\partial_t\vert z}
\label{eq19}
\ee
From Eq.\,\pref{eq14}
\be
-i\braket{\bar z\vert\partial_t\vert z}=-i\bar z(t)\partial_t z(t)=
\dot\phi\frac{1-\cos\theta}{2}=\frac{1}{2}{\dot{\Omega}}
\ee
where $\dot\Omega$ is the rate at which the vector $\hat n$ sweeps out
solid angle in channel space. 
The total accumulated Berry phase
associated with a closed path in channel space is then 
\begin{equation}\label{}
\eta = \frac{1}{2}\int dt \dot\Omega= \frac{1}{2}\Omega
\end{equation}
where $\Omega$ is the total solid angle subtended by the path. 
This has the general form of $\eta=\Omega S$. The fact that we find
the pre-factor $S=1/2$ shows the spinorial character of the ground
state and the underlying  order parameter.

\subsection{Strong Coupling analysis of the single-channel Kondo Model}\label{}

\begin{figure}[h!]
\includegraphics[width=\linewidth]{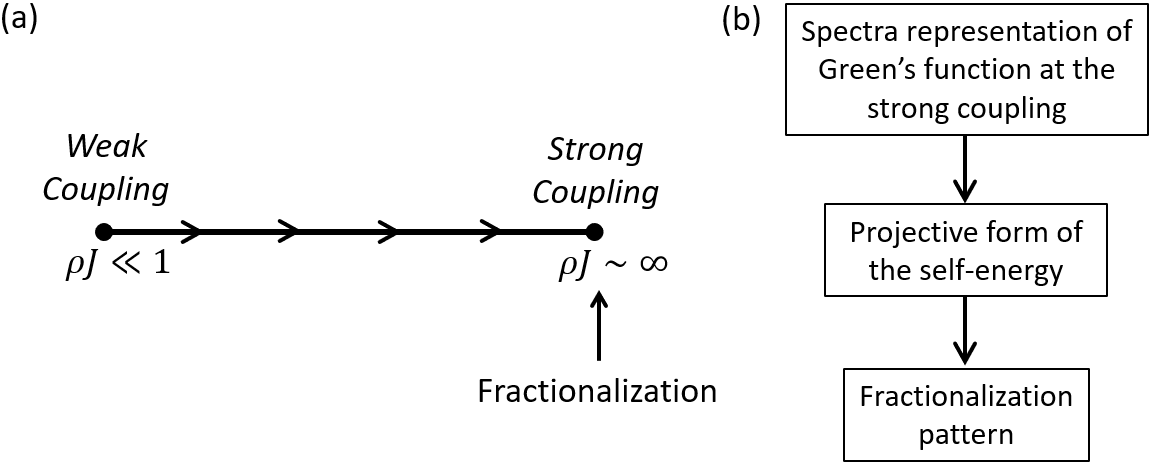}
\caption{\small (a) The renormalization group flow for the single-channel Kondo problem. Fractionalization is a property of the strong coupling fixed point. (b) The logic followed in this section. We first use the Lehman representation to find the Green's function at strong coupling. Then use the latter to extract the self-energy which has a projective form. And finally, having obtained the self-energy we can read off the fractionalization pattern from the strong coupling Hamiltonian.}\label{figscheme1}
\end{figure}

The goal of this supplementary material is to display the patterns of order fractionalization
that occur when symmetry breaking terms are added to the strong
coupling one and two channel Kondo models. The basic 
idea is to first, recognize that the fractionalization is a property of the strong coupling fixed point (Fig.\,\ref{figscheme1}a) and can be studied by strong coupling expansion. This is schematically shown in Fig.\,\ref{figscheme1}(b). 

We begin with the one channel model, demonstrating that self-energy has a sharp pole $\Sigma (z)= V^{2}/z$, that the composite operator $(\vec{ \sigma }\cdot \vec{S})_{\alpha \beta }\psi_{\beta }\equiv V f_{\alpha }$. Next, we consider the an extension of the single-channel Kondo model known as the compactified two-channel Kondo model \cite{colemanioffe}, in which the conduction sea has an SO(4)$\sim$SU(2)$\times$SU(2) symmetry.  By breaking 
this symmetry down to an SO(3) subgroup, we are able to demonstrate the fractionalization of the spin into Majorana fermions, giving rise to an odd-frequency triplet state. 

\subsubsection{Fractionalization in the strong coupling limit of single-channel Kondo
model}\label{}

The strong coupling Kondo lattice is literally just the interaction
term in the Kondo model
\begin{equation}\label{}
H = J 
( \psi\dg \vec{\sigma }\psi)\cdot \vec{S}
\end{equation}
where we use the notation $\psi\dg \equiv (\psi\dg_{\uparrow},\psi\dg_{\dw })$.
This is a trivial problem, but we can gain some interesting insight
into the nature of composite fermions.  We can write this Hamiltonian
as $H/J =  \left(\vec{S}_{\psi}+ \vec{S} \right)^{2}- (\vec{
S}_{\psi})^{2}-\frac{3}{4}$, with eigenvalues $E= -3J/2$, $E=J/2$ for the
singlet and triplet respectively, and $E=0$ for the empty and doubly
occupied conduction electron states. 
The Hilbert space of energy eigenstates 
$\{ \vert \lambda\rangle  \}$ contains the ground-state
\begin{eqnarray}\label{l}
\vert \phi \rangle  = 
\frac{1}{\sqrt{2}}\left(
\psi\dg_{\downarrow}\vert
\Uparrow \rangle - \psi\dg_{\uparrow}\vert  \Downarrow\rangle 
\right)
, \qquad
E_{\phi  }= - \frac{3}{2}J
\end{eqnarray}
the empty and doublet occupied states, 
\begin{equation}\label{}
\left. 
\begin{array}{rcl}
\vert \sigma \rangle &=& \{\vert \Uparrow\rangle, \vert \Downarrow
\rangle\},\cr
\vert 2,\sigma \rangle&=& \{\vert 2,\Uparrow\rangle, \vert 2,\Downarrow\rangle \}= \{  \psi\dg_{\uparrow }\psi\dg_{\downarrow}
\vert \Uparrow \rangle,
\psi\dg_{\uparrow }\psi\dg_{\downarrow}\vert \Downarrow\rangle\}, 
\end{array}
\right\}
 \qquad  E_{\sigma } = 0
\end{equation}
and the triplet states
\begin{equation}\label{}
\left.\begin{array}{rcl}
\vert t, m_{J}=+1\rangle &=&\psi\dg_{\uparrow}\vert \Uparrow\rangle \cr
\vert t,m_{J}=0\rangle &=&\frac{1}{\sqrt{2}}
\left(
\psi\dg_{\uparrow}\vert
\Downarrow \rangle + \psi\dg_{\downarrow}\vert  \Uparrow\rangle 
 \right)\cr
\vert t, m_{J}=-1\rangle &=&\psi\dg_{\downarrow }\vert \Downarrow \rangle 
\end{array} \right\}, \qquad E_{t}= \frac{J}{2}
\end{equation}
We can use the Lehmann representation to compute the electron Green's
function, writing
\begin{eqnarray}\label{l}
G_{c} (z) 
&=& \sum_{\lambda} 
\left[\frac{|\langle \lambda\vert \psi\dg_{\sigma
}\vert \phi \rangle |^{2}}{z - (E_{\lambda }- E_{\phi })} + 
\frac{|\langle \lambda\vert \psi_{\bar \sigma
}\vert \phi \rangle |^{2}}{z - (E_{\phi })-E_{\lambda}} \right]\cr
&=& 
\left[\frac{|\langle 2,\sigma \vert \psi\dg_{\sigma
}\vert \phi \rangle |^{2}}{z - (E_{\lambda }- E_{\phi })} + 
\frac{|\langle \sigma \vert \psi_{\bar \sigma
}\vert \phi \rangle |^{2}}{z - (E_{\phi })-E_{\lambda}} \right].
\end{eqnarray}
where the simplification occurs because the creation and annihilation
operators only link the ground-state with the doubly occupied and
empty states, respectively.  We can use the matrix elements
\begin{eqnarray}\label{l}
\langle 2,\sigma \vert \psi\dg_{\sigma '}\vert \phi \rangle &= &\frac{1}{\sqrt{2}}\delta_{\sigma \sigma '},\cr
\langle \sigma \vert \psi_{\bar  \sigma '}\vert \phi \rangle &= &-\frac{1}{\sqrt{2}}\delta_{\sigma \sigma '}.
\end{eqnarray}
to obtain
\begin{equation}\label{}
G_{c} (z) = \frac{1}{2}\left[\frac{1}{z-\frac{3J}{2}} +
\frac{1}{z+\frac{3J}{2}}\right] = \frac{z}{z^{2}- (3J/2)^{2}}= 
\frac{1}{z - \Sigma (z)}
\end{equation}
where
\begin{equation}\label{}
\Sigma (z)= \frac{(3J/2)^{2}}{z}\equiv \frac{V^{2}}{z}.
\end{equation}
thus demonstrating the presence of a sharp pole in the electron self
energy with ``hybridization'' $V=3J/2$. The form of this self-energy
is consistent with the identification
\begin{equation}\label{}
J(\vec{\sigma }_{\alpha \beta }\cdot
\contraction{}
{\vec{S}}
{)}
{c }
\vec{S})
\psi_{\beta } =  V \hat f_{\alpha }.
\end{equation}
where $V=3J/2$. This strong coupling analysis can be extended in two ways which we leave for a separate publication and just mention the result: a) by aplying a magnetic field to the spin, it can be shown that the action of the spin operator $\vec{S}$ on the ground-state is described by $\vec{S}\equiv f\dg
\frac{\vec{\sigma }}{2}f$. b) by perturbatively including the effect of the kinetic Hamiltinian of the conduction band, it can be shown that the $c$ and the emergent $f$-electrons have a simple two-band effective Hamiltonian at low-energies. 

\subsubsection{Majorana Fractionalization }\label{}

Next, we look at the strong coupling regime of the so-called compactified two-channel Kondo problem \cite{colemanioffe}. 
\be
H_{int}=J\tpsi\dg[\vec\sigma+(1-\eps)\vec\tau] \tpsi\cdot \vec S, \qquad\qquad \tpsi\dg\equiv\matc{cccc}{\psi\dg_\ua & \psi\dg_\da & \psi\dn_\da & -\psi\dn_\ua}
\ee
Here, $\tpsi\dg\vec\sigma\tpsi$ and $\tpsi\dg\vec\tau\tpsi$ are the spin and isospin densities of the conduction electrons and
\be
\frac{1}{2}\tpsi\dg\vec\tau\tpsi\cdot S=(\psi\dg_\ua\psi\dn_\ua+\psi\dg_\da\psi\dn_\da)S^z+(\psi\dg_\ua\psi\dg_\da S^-+h.c.)
\ee
For $\eps\to 1$ this is just a perturbation of the single-channel Kondo impurity: Near the strong coupling fixed point, the second term is irrelevant because the site near the spin impurity cannot contain both spin and isospin. {However, there is a quantum critical point \cite{Bulla97} at $\eps=0$ analogous to the two-channel Kondo problem, which happens at strong coupling $J\to\infty$.}

\begin{figure}[h!]
\includegraphics[width=1\linewidth]{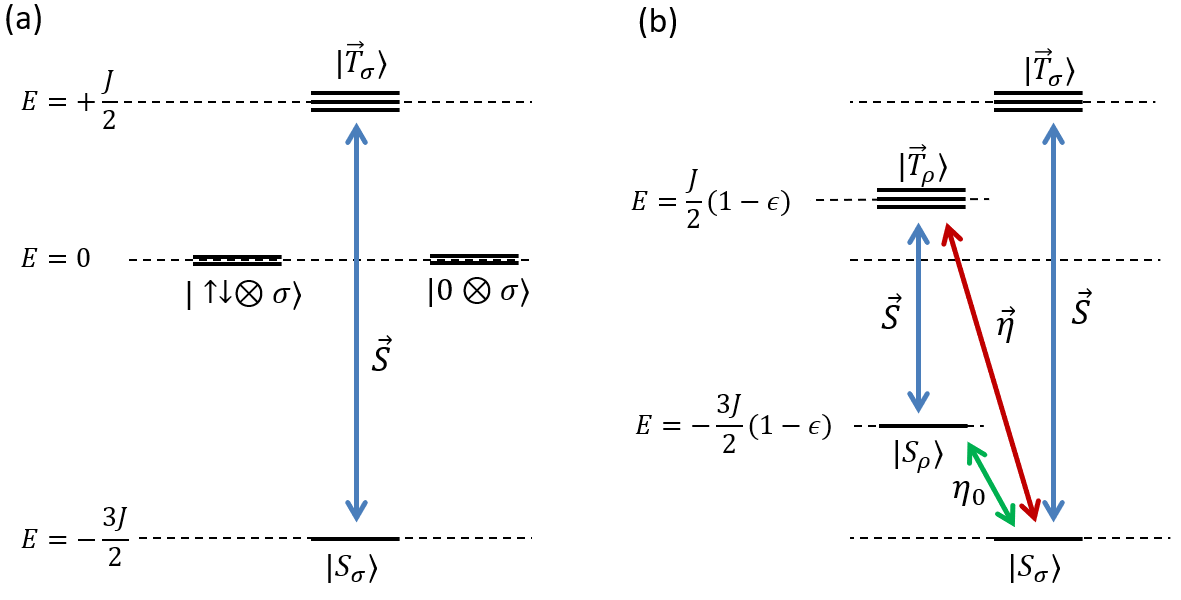}
\caption{\small The energy diagram for a single-site problems. (a) Single-channel Kondo problem (b) The compactified two-channel Kondo problem. The arrows show the effect of acting with various operators on the ground state.\label{figdiag}}
\end{figure}

The local Hilbert space of the single-channel Kondo problem is composed of the 8 states. In the usual Kondo problem, the Kondo interaction mixes spin states into spin-singlet $\ket{S_\sigma}$  and spin-triplets $\ket{\vec T_\sigma}$ separated by an energy gap of $2J$, while the doubly occupied and empty states remain at zero energy [Fig.\,\ref{figdiag}(a)]. In the comapctifie 2CK model, the empty and doubly occipied states are also mixed to create the charge-singlet and charge-triplets states:
\be
\ket{S_\rho}=\frac{\ket{\Ua 0}-\ket{\Da \ua\da}}{\sqrt{2}}, \qquad \ket{T^0_\rho}=\frac{\ket{\Ua 0}+\ket{\Da \ua\da}}{\sqrt{2}}, \qquad \ket{T^+_\rho}=\ket{\Ua \ua\da}, \qquad \ket{T^-_\rho}=\ket{\Da 0}\nonumber
\ee
as shown in Fig.\,\ref{figdiag}(b). This problem is best treated in the Majorana representation. We can introduce four Majorana fermions $\gamma_0$ and $\vec\gamma$ via
\be
\mat{\psi_\ua \\ \psi_\da}=\frac{1}{\sqrt 2}[\gamma_0-i\vec\sigma\cdot\vec\gamma]\mat{0 \\ i}=\frac{1}{\sqrt 2}\mat{\gamma^x-i\gamma^y \\ -\gamma^z+i\gamma^0}
\ee
$\gamma_0$ and $\vec\gamma$ are conduction band Majorana fermions that transform as a scalar and vector under SU(2) transformation. There is a freedom on assigning Majorana-s to the Dirac (i.e. complex) fermions, which is reflected by the choice of the spinor on the right. For this choice, we have
\be
\tpsi\dg\vec\sigma\tpsi=i(-\gamma_0\vec\gamma-\frac{1}{2}\vec\gamma\times\vec\gamma),\qquad \tpsi\dg\vec\tau\tpsi=i(\gamma_0\vec\gamma-\frac{1}{2}\vec\gamma\times\vec\gamma)
\ee
which leads to
\be
H_{int}=-iJ[\eps\gamma_0\vec\gamma+\frac{1}{2}(2-\eps)\vec\gamma\times\vec\gamma]\cdot\vec S\label{eqHamM}
\ee
Using the zero-temperature spectral representation for the Green's function
\be
G_{\alpha\beta}(z)=\sum_\lambda \Big[\frac{\braket{g\vert \gamma_\alpha\vert\lambda}\braket{\lambda\vert\gamma_\beta\vert g}}{z-(E_\lambda-E_g)}+\frac{\braket{\lambda\vert \gamma_\alpha\vert g}\braket{g\vert\gamma_\beta\vert \lambda}}{z+(E_\lambda-E_g)}\Big]
\ee
and the matrix elements indicated in Fig.\,\ref{figdiag}(b) find
\be
G_{\alpha\beta}(z)=\frac{P^S_{\alpha\beta}}{z-\Delta E_S/z}+\frac{P^T_{\alpha\beta}}{z-\Delta E_T/z}
\ee
where in this Majorana basis we have $P^S_{\alpha\beta}=\delta_{\alpha\beta}\delta_{\alpha,0}$ and $P^T_{\alpha\beta}=\delta_{\alpha\beta}-P^S_{\alpha\beta}$, and 
\be
\Delta E_S=\frac{3J}{2}\eps, \qquad \Delta E_T=\frac{J}{2}(4-\eps).
\ee
This means we can write the self-energy as
\be
\Sigma_{\alpha\beta}(z)=P_{\alpha\beta}^S\frac{\Delta E_S}{z}+P_{\alpha\beta}^T\frac{\Delta E_T}{z}
\ee
Note that in the limit of $\eps\to 1$ this expression reduces to the original single-channel self-energy. On the other-hand in the limit of $\eps\to 0$ the singlet contribution drops out.

Having the structure of the self-energy, we can work out the fractionalization pattern by which the conduction electron Majoranas combine with the spin. The contractions allowed by the symmetry are
\be
\contraction[1ex]{}{\gamma_0}{}{S^a}
\gamma_0 S^a\to A \eta^a, 
\qquad 
\contraction[1ex]{}{\gamma^a}{}{S^b}
\gamma^a S^b\to \delta^{ab}B\eta_0+\eps^{abc}C\eta^c.
\ee
in terms of emergent localized Majorana fermions $\eta_0$ and $\vec\eta$. Using these in Eq.\,\pref{eqHamM} we find
\be
H_{int}\to -iJ\Big\{\Big[2C(2-\eps)-A\eps\Big]\vec\gamma\cdot \vec\eta+3B\eps\gamma_0\eta_0\Big\}
\ee
and comparing it to the projective self-energy, the coefficients can be read off as 
\be
B=\frac{1}{2}, \qquad A=-\frac{1}{2},\qquad C=\frac{1}{2}.
\ee
Especially, at the $\eps\to 0$ we recover the fractionalization pattern reported in the entry a' of the table in the main section.

\subsection{Analysis of SO(5) symmetry breaking in the two-channel Kondo model
}\label{}

\begin{figure}[h!]
\includegraphics[width=0.8\linewidth]{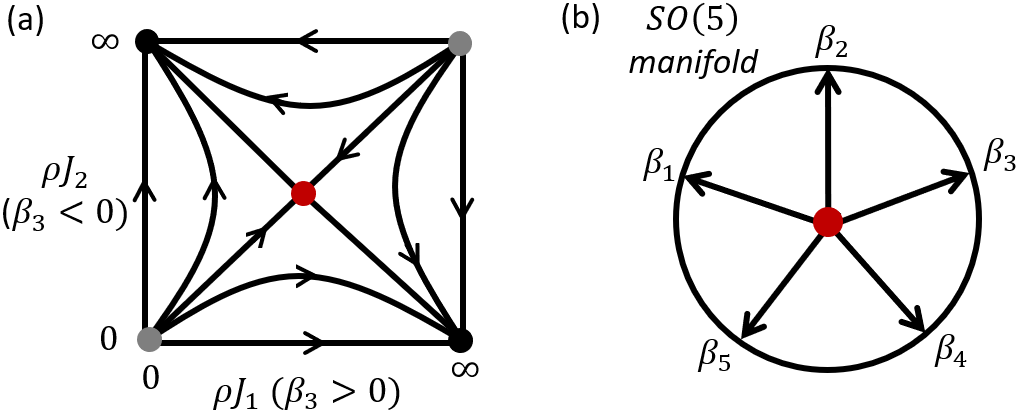}
\caption{\small (a) The renormalization group flow for the two-channel Kondo problem in the space of Kondo couplings of the two channel. The critical fixed point (red) is sensitive to the channel asymmetry and flows to stable fixed points (black). (b) More general representation of the symmetry breaking; the critical finite-coupling fixed point of the two-channel Kondo lattice can be (explicitly or spontaneously) broken into any point on the SO(5) manifold.}\label{figscheme2}
\end{figure}

Finally, we turn to the usual two channel Kondo model (Fig.\,\ref{figscheme2}) which has a continuous channel-charge SP (4) and spin SU(2) symmetry, SP(4)$\times$ SU(2).  

In the paper we used NRG to induce order fractionalization in an impurity with channel asymmetry
\be
H=\sum_\alpha\psi\dg_\alpha (\vec\sigma\cdot\vec S)[J+\delta J\alpha^z]\psi_\alpha.\label{eqalz}
\ee
At the strong coupling, the self-energy is only induced in the stronger channel. This is consistent with the the RG flow (Fig.\,\ref{figscheme2}a) which imply that $J,\delta J\to\infty$ and $\delta J/J\to 1$ so that the Hamiltonian takes the same projective form as the self-energy.

We also argued that since the channel symmetry is a continuous group, we may induce order fractionalization using the relevant perturbation
\be
H=\sum_\alpha\psi\dg_\alpha (\vec\sigma\cdot\vec S)[J+\delta J\vec n\cdot\vec\alpha]\psi_\alpha.\label{eqalvec}
\ee
At the strong coupling fixed point the self-energy is only induced in the stronger channel, leading to a projective form for the self-energy
\be
\Sigma_{\lambda\lambda'}(z)=V^2\frac{P_{\lambda\lambda}}{z}, \qquad P=\frac{1}{2}(1+\vec n\cdot\vec\alpha)
\ee
and the fractionalization pattern 
\be
J(\vec\sigma\cdot\vec S)\psi_\alpha=V_\lambda f,\label{eq35}
\ee
where $\psi_\alpha$ with $\alpha=1,2$ and $f$ are regarded as spinors in spin-space and $(\vec n\cdot\alpha)V_\lambda=+V_\lambda$. 

In this section, we first review the symmetry properties of the two-channel Kondo model, showing that the channel symmetry is part of a larger channel-charge SP(4) symmetry. The fully symmetry of the two-channel Kondo problem is SO(3)$\times$ SO(5) or SU(2)$\times$SP(4) \cite{Affleck92}. It turns out that there is a full family of projection operators, which enables us to describe a family of different fractionalization patterns (Fig.\,\ref{figscheme2}b).

\subsubsection{SP(4) symmetry of the two-channel Kondo problem}
In order to see this, note that the Kondo interaction can be written as \cite{Affleck88}
\be
H_{int}=\frac{J}{4}\tr{\Psi\dg\Psi \vec\sigma^T\cdot\vec S}
\ee
where we have arranged the elements of $\tilde\psi$ in a 4$\times$2 matrix
\be
\Psi_{\lambda,\sigma}=\matc{c|c}{
\psi_{1\ua} & \psi_{1\da} \\
\psi_{2\ua} & \psi_{2\da} \\
\hline
-\psi\dg_{2\da} & \psi\dg_{2\ua}\\
\psi\dg_{1\da} & -\psi\dg_{1\ua}}\label{eq29}
\ee
Despite its formidable form, different symmetries of the problem are manifest in this representation. 
The quantum number $\sigma=\ua,\da$ is connected to the column, whereas the quantum number $\lambda=[\alpha,\tau]$ connected to the rows is a super-index characterizing channel $\alpha=1,2$ and the isospin $\tau=e,h$ degrees of freedom. The isospin tags the large blocks whereas channel tags internal structure of the blocks. Associated with each of these quantum numbers, there exist Pauli matrices $\vec\sigma$ acting from the right side and Pauli matrices $\vec\alpha$ and $\vec\tau$ acting from the left side. The Kondo interaction, as well as the commutation relations and the kinetic energy are invariant under separate left and right transformations of $\psi$:
\be
\Psi\to g\Psi, \qquad \Psi\to \Psi h\label{eq28}
\ee
where $h=e^{i\vec\phi\cdot\vec\sigma}$ is a usual SU(2) transformation, but $g$ is an element of SP(4) parametrized by
\be
{\rm SP(4)}:\qquad g=e^{i\sum_{m=1}^{10}\theta_j\Lambda_j}
,\qquad \Lambda_j\in\{\vec\alpha,\vec\alpha\tau_1,\vec\alpha\tau_2,\tau_3\}.\label{eqgen}
\ee
\emph{Proof:} {\small  Naively, for any SU(4) and SU(2) transformation can be used in Eq.\,\pref{eq28}. However, our description Eq.\,\pref{eq29} has a redundancy, that implies invariance under the discrete transformation
\be
\tilde\psi\to \theta_L\tilde\psi^*\theta_R, \qquad \theta_L=i\alpha_2\tau_1, \qquad \theta_R=i\sigma_2\label{eq29}
\ee
that needs to remain consistent under unitary transformations. For the left transformation, this means
\be
g\theta_L\tilde\psi^*\theta_R=\theta_L(g\tilde\psi)^*\theta_R, \so g\theta_L=\theta_Lg^*
\ee
The SU(4) transformation can be generated by the generators $\Lambda_{jk}=\alpha_j\tau_k$.
Considering a group element $g$ to $g=e^{i\Lambda}$ in terms of generators $\Lambda=\sum_{jk}d_{jk}\Lambda_{jk}$, with infinitesimal arbitrary $d_{jk}$, we conclude
\be
\Lambda=\theta_L\Lambda^T\theta_L
\ee
which is the defining equation for symplectic representation SP(4). The subset of generators that satisfy this condition are the 10 generators given in Eq.\,\pref{eqgen}. This is the same number of generators as SO(5) and indeed it can be shown that SP(4) id a double cover of SO(5). The discrete symmetry \pref{eq29} is not restrictive for the right-transformation, because it requires
\be
\theta_Rh=h^*\theta_R
\ee
which in terms of infinitesimal group element $g=e^{i\Gamma}$ with $\Gamma=\sum_jd_j\sigma_j$ leads to 
\be
\Gamma^T\theta_R+\Gamma\theta_R=0
\ee
All Pauli matrices satisfy this condition and so the $h$-transformations remain SU(2).}
\subsubsection{The projective form of the self-energy}
The 10 generators $\Lambda_j\in$ SP(4) are a subset of the 15 generators of SU(4). Out of the remaining 5 generators that are not part of $g\in SP(4)$ we can build a family of projectors
\be
{\cal P}=\frac{1}{2}\Big(1+\vec m\cdot\vec\beta\Big), \qquad \vec\beta=\{\alpha_1\tau_3,\alpha_2\tau_3,\alpha_3\tau_3,\tau_1,\tau_2\},\label{eqp}
\ee
parametrized by the 5 component vector $\vec m$. The important point is that these projectors rotate among themselves by the elements of $g\in$SP(4) group:
\be
g\dg (\vec m\cdot\vec\beta) g=\vec m'\cdot\vec\beta.
\ee
The channel symmetry breaking perturbations discussed in Eqs.\,\ref{eqalz} and \ref{eqalvec} correspond to $\alpha_3\tau_3$ and $\vec\alpha\tau_3$, respectively. Since, different projectors can be rotated into each other, for the most general case of Eq.\,\ref{eqp} the self-energy has the projective form
\be
\Sigma(z)=\frac{V^2}{z}{\cal P}.
\ee
in the $\Psi$ representation. To extract the patterns of fractionalization, we first find eigenvectors of $\cal P$, i.e. look for unit vectors $z$ that satisfy
\be
(\vec m\cdot\vec\beta){\rm z}={\rm z}
\ee
with $z\dg z=1$. Such an eigenvector enables a CP$^3$ representation of the five-component $\vec m$ vector through $\vec m={\rm z}\dg \vec\beta {\rm z}$. These eigenvectors have the general form
\be
{\rm z}=\mat{u_1 \\ u_2 \\ -\bar v_2 \\ v_2}.
\ee
It turns out that ${\cal P}$ has a symmetry
\be
\eps_4^T(\vec m\cdot\vec\beta)\eps_4=\vec m\cdot\vec \beta
\ee
where $\eps_4=i\alpha_2\tau_1$ is the fully antisymmetric matrix. It follows that both ${\rm z}$ and $\eps_4{\rm z}^*$ are degenerate eigenvectors of ${\cal P}$ with eigenvalue +1:
\be
(\vec m\cdot\vec\beta)(\eps_4{\rm z}^*)=(\eps_4{\rm z}^*).
\ee
These two can be combined into a single matrix
\be
{\cal Z}_{\lambda\tau}=\mat{{\rm z} & \eps_4{\rm z}^*}=\matc{c|c}{u_1 & v_1 \\ u_2 & v_2 \\ \hline -\bar v_2 & \bar u_2 \\ \bar v_1 & -\bar u_1}=g\matc{c|c}{1 & 0 \\ 0 & 0 \\ \hline 0 & 0 \\ 0 & 1}\label{eqz}
\ee
Here, $\lambda$ index corresponds to the rows and the eigenvectors are enumerated by the coulomn $\tau$ index. The second equation highlights that the general ${\cal Z}$ eigen-matrix can be obtained form a SP(4) rotation of a reference ${\cal Z}$ eigen-matrix, corresponding to the projector $\beta=\alpha_3\tau_3$ used in the NRG.

Writing ${\rm z}=\braket{\lambda\vert \tau=1}$ and $\eps_4{\rm z}^*=\braket{\lambda\vert \tau=2}$ the projector has the form ${\cal P}=\sum_{\tau=1,2}\ket{\tau}\bra{\tau}$ or in components:
\be
{\cal P}_{\lambda\lambda'}=\sum_{\tau=1,2}\braket{\lambda\vert\tau}\braket{\tau\vert\lambda'}={\cal Z}\dn_{\lambda\tau}{\cal Z}^*_{\lambda'\tau}.
\ee
and the self-energy will take the form
\be
\Sigma_{\lambda\lambda'}(z)=\frac{V^2}{z}({\cal Z}{\cal Z}\dg)_{\lambda\lambda'}
\ee
and has an internal SU(2) gauge invariance under ${\cal Z}\to {\cal Z}h$ where $h\in SU(2)$.

\subsubsection{Fractionalization pattern and composite ordering}
In terms of $\Psi$ matrix the Hamiltonian is
\be
H=\frac{J}{4}\tr{\Psi\dg {\cal P}\Psi (\vec\sigma^T\cdot S)}.
\ee
and Eq.\,\pref{eq35} can be written as
\be
J\Psi (\vec\sigma^T\cdot\vec S)={\cal V}{\cal F}\label{eqfub}, \qqquad \tilde {\cal F}=\mat{f_\ua & f_\da \\ 	f\dg_\da & -f\dg_\ua},
\ee
where ${\cal V}=V{\cal Z}$. Using the notation $Vu_\lambda\to V_\lambda$ and $Vv_\lambda \to \Delta_\lambda$ in the general expression for $\cal Z$ in Eq. (\ref{eqz}) and translating the above equation back to the $\psi$ spinors, the contraction reads
\be
J(\vec\sigma\cdot \vec S)_{\alpha\beta}\psi_{\lambda\sigma}=V_\lambda f_\alpha+\Delta_\lambda f\dg_{-\alpha}\sgn(\alpha)
\ee
which is nothing but the composite pairing \cite{Kuramoto}. In particular the conduction electron self-energy contains Andreev reflections. This is not surprising given that here, we have induced the symmetry breaking in the pairing channel. As explained in the manuscript, we conjecture that this symmetry breaking and the associated fractionalizations can happen spontaneously in a lattice.